\documentclass[pdflatex,sn-mathphys-num]{sn-jnl}

\usepackage{graphicx}%
\usepackage{multirow}%
\usepackage{amsmath,amssymb,amsfonts}%
\usepackage{amsthm}%
\usepackage{mathrsfs}%
\usepackage[title]{appendix}%
\usepackage{xcolor}%
\usepackage{textcomp}%
\usepackage{manyfoot}%
\usepackage{booktabs}%
\usepackage{algorithm}%
\usepackage{algorithmicx}%
\usepackage{algpseudocode}%
\usepackage{listings}%

\theoremstyle{thmstyleone}%
\theoremstyle{thmstyletwo}%

\theoremstyle{thmstylethree}%

\begin{document}

\title[Validating LLM-generated political networks with survey-reported partisan exposure]{Validating LLM-generated political networks with survey-reported partisan exposure}


\author*[1,2]{\fnm{Marco} \sur{Tonin}}\email{marco.tonin@unitn.it}

\author[1,2]{\fnm{Michele} \sur{Tizzoni}}\email{michele.tizzoni@unitn.it}

\author[2]{\fnm{Bruno} \sur{Lepri}}\email{lepri@fbk.eu}

\author[3]{\fnm{Luca Maria} \sur{Aiello}}\email{luai@itu.dk}

\affil[1]{\orgdiv{Department of Sociology and Social Research}, \orgname{University of Trento}, \orgaddress{\street{Via Verdi 26}, \city{Trento}, \postcode{I-38122}, \country{Italy}}}

\affil[2]{\orgdiv{Mobile and Social Computing Lab (MobS)}, \orgname{Fondazione Bruno Kessler}, \orgaddress{\street{Via Sommarive 18}, \city{Trento}, \postcode{I-38123}, \country{Italy}}}

\affil[3]{\orgdiv{Data Science Section}, \orgname{IT University of Copenhagen}, \orgaddress{\street{Rued Langgaards Vej 7}, \city{Copenhagen}, \postcode{DK-2300}, \country{Denmark}}}


\abstract{Large language models (LLMs) are increasingly used to simulate human behavior in social science research, including decisions, interactions, and social relationships. 
However, in network formation tasks, existing evaluations often focus on aggregate network properties, while networks with plausible structural properties may still fail to reproduce individuals' social and political environments.
Here, we generate LLM-based political networks using personas sampled from the nationally representative ANES 2020--2022 Social Media Study and use respondents' survey-reported partisan exposure as an empirical benchmark for the political environments generated for the corresponding personas.
We first characterize the structural properties and political assortativity of the generated networks, and then compare ego-level partisan exposure with the survey.
We find that the structure and political assortativity of the generated networks vary across experimental configurations.
More importantly, when comparing generated partisan exposure with the survey, some models generate stronger separation between Democratic and Republican egos than reported in the survey, with higher same-party and lower cross-party exposure, while others mainly underestimate Republican exposure across ego groups.
Moreover, networks with similar political assortativity can still generate different partisan environments around the same personas, including among networks with plausible structural properties.
Finally, generated networks show group-specific errors, under- or overestimating exposure to either Democrats or Republicans for specific social groups. 
These findings show that evaluating LLM-based social simulations only through aggregate outcomes can hide errors in the local environments through which those outcomes emerge, highlighting the importance of validation at both aggregate and individual levels.}


\maketitle


\section{Introduction}\label{sec1}


Large language models (LLMs) are increasingly used as tools for social simulation because they can generate human behavioral traces conditioned on attributes and beliefs that can be experimentally manipulated~\cite{bail_can_2024,lu_llms_2024,broska_mixed_2025}.
Recent work uses LLMs to generate survey responses, simulate individual decisions, model political attitudes, and represent interactions among synthetic or survey-derived personas~\cite{argyle_out_2023,aher_using_2023,park_generative_2023,cheng_compost_2023,broska_mixed_2025,kolluri_finetuning_2025,romberg_hybrid_2026}.
Yet the validity of these simulations remains an open challenge~\cite{cheng_compost_2023,rossi_problems_2024,bisbee_synthetic_2024,wang_large_2025,larooij_validation_2025,krsteski_valid_2026,
barrie_synthetic_2026,munker_dont_2026}.

Simulations of heterogeneous human populations rely largely on priming the model with \emph{personas} that specify sets of demographic or political characteristics~\cite{argyle_out_2023, cheng_compost_2023, lee_spectrum_2025}.
LLMs, however, may reduce multidimensional personas to simplified demographic profiles, flattening heterogeneity and reproducing stereotypes or group-specific biases~\cite{santurkar_whose_2023, lee_large_2024, gupta_bias_2024, hu_generative_2024, wang_large_2025}.
These distortions may compound when LLMs are used not only to generate individual responses but also collective dynamics~\cite{ashery_emergent_2025, flint_group_2026} or networks of social relationships~\cite{chang_llms_2025, papachristou_network_2024, piao_emergence_2025, kilaru_when_2026}, which are shaped by individual characteristics and by patterns of homophily and assortative mixing~\cite{mcpherson_birds_2001, kossinets_origins_2009, rivera_dynamics_2010, smith_social_2014}. 
These concerns become particularly relevant, for example, when LLMs are used to simulate patterns of political polarization in social networks~\cite{chang_llms_2025,piao_emergence_2025}, given that people's political stances are influenced by the ideological composition of their social contacts and political environments~\cite{huckfeldt_citizens_1995,sinclair_social_2012,campbell_social_2013,huckfeldt_networks_2014,enos_causal_2014,enos_space_2017,butters_polarized_2022,tonin_physical_2025}.

Crucially, however, these distortions may go unnoticed because models can produce realistic collective outcomes in aggregate while deviating from realistic human behavioral traces at the level of subgroups or individuals~\cite{arruda_collective_2026}.
A synthetic network generated by LLM agents with political personas, for example, may display realistic density, clustering, modularity, assortativity, and political homophily, yet still fail to reproduce the distribution of political contacts observed in real-world data for individuals or specific political subgroups.
Existing evaluations of LLM-generated social networks nonetheless focus on coarse structural properties such as density or modularity~\cite{chang_llms_2025,kilaru_when_2026,papachristou_network_2024,de_marzo_emergence_2023}. 
This limited analytic resolution stems largely from the scarcity of fine-grained empirical data that would allow researchers to assess how well synthetic networks reproduce individual nodes' perspectives on their surrounding social environments.

In this paper, we use a persona-based approach to generate LLM-based social networks grounded in real-world survey data from the American National Election Studies (ANES) 2020-2022 Social Media Study~\cite{anes_socialmedia_2023}, which includes political attributes and self-reported exposure to Democratic and Republican social contacts, and use respondents' reported partisan exposure as an empirical benchmark for the political environments generated around the corresponding personas.
First, we test whether the resulting networks exhibit plausible high-level structural properties.
Second, we assess whether LLM-generated ties reproduce survey-reported patterns of individual-level partisan exposure, comparing simulated ego networks to survey-based measures.
Third, we examine whether networks with similar political assortativity generate similar partisan environments around the same personas, and whether exposure errors vary across demographic and political groups.

We find that both the structure and political assortativity of LLM-generated networks vary across models and prompting configurations.
More importantly, comparison with survey-reported partisan exposure reveals differences across models and configurations in both the direction and magnitude of the errors.
Specifically, some models generate stronger separation between Democrats and Republicans than reported in the survey, with higher same-party and lower cross-party exposure, while others mainly underestimate Republican exposure across ego groups.
Moreover, networks with similar political assortativity can still generate different partisan environments around the same personas, including among networks with plausible structural properties.
Finally, exposure errors vary across demographic and political groups, and across models.

Overall, these results show that LLM-generated political networks differ across models and prompting configurations in both their aggregate properties and the partisan environments generated at the individual level. 
In this context, survey-reported exposure provides an empirical benchmark to evaluate whether the generated networks reproduce the political environments experienced by respondents beyond what can be captured by network-level measures alone.

\begin{figure}[t!]
\centering
\includegraphics[width=1\textwidth]{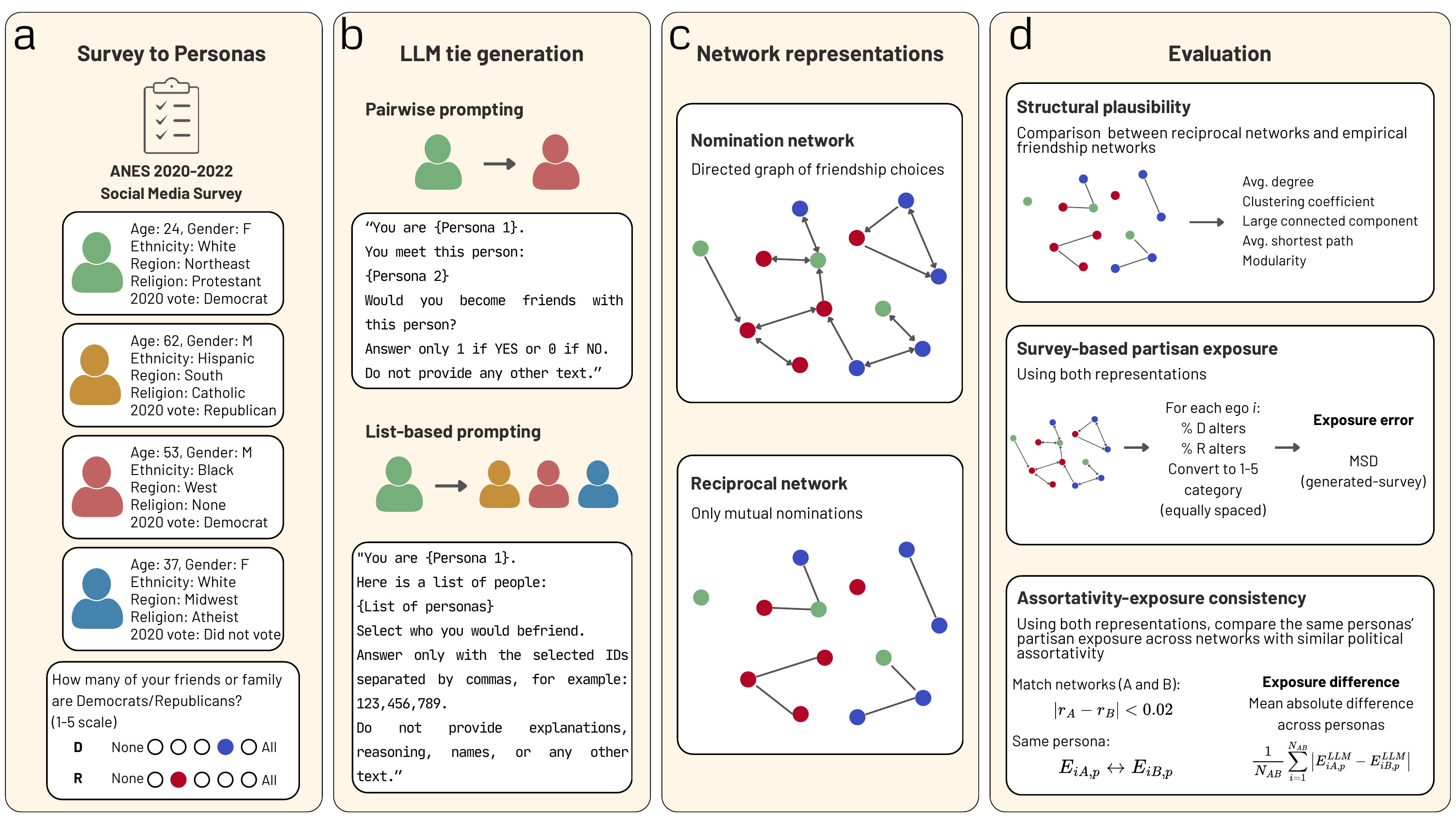}
\caption{\textbf{Workflow for generating and evaluating LLM-based social networks.} \textbf{(a)} We sample respondents from the ANES 2020--2022 Social Media Study and construct personas using their demographic and political attributes. The survey also provides self-reported exposure to Democratic and Republican friends or family on a 1--5 scale, which is used as an empirical benchmark.
\textbf{(b)} We generate friendship nominations using pairwise and list-based prompting methods, asking each persona to select potential friends.
\textbf{(c)} From these choices, we construct two network representations: a directed nomination network containing all ego-to-alter friendship choices, and an undirected reciprocal network containing only mutually selected ties.
\textbf{(d)} We evaluate the generated networks along three dimensions: structural plausibility, by comparing reciprocal networks with empirical friendship-network benchmarks; survey-based partisan exposure, by comparing generated ego-level exposure with the corresponding ANES measure; and assortativity--exposure consistency, by assessing whether networks with similar political assortativity generate similar partisan exposure for the same personas.
}
\label{fig1}
\end{figure}

\section{Results}\label{sec2}

Using a persona-based approach (see Methods and Fig.~\ref{fig1}), we generate social networks from respondents in the ANES 2020-2022 Social Media Study across five models, temperatures, attribute orderings, and different prompting configurations, representing friendship nominations between the same 100 survey-based personas.
We evaluate the generated networks based on their structural properties, the partisan exposure they generate around individual personas, and the extent to which aggregate assortativity captures these local political environments.
We first examine how network structure and assortativity vary across models and prompting configurations, and to what extent generated networks appear structurally plausible.
Second, we compare ego-level partisan exposure in the generated networks with exposure reported in the ANES 2020-2022 Social Media Survey.
Finally, we test whether networks with similar levels of political assortativity also generate similar partisan environments for the same individuals, and assess how exposure errors vary across demographic and political groups.

\subsection{LLM-generated network structure varies across configurations}

\begin{figure}[t!]
\centering
\includegraphics[width=1\textwidth]{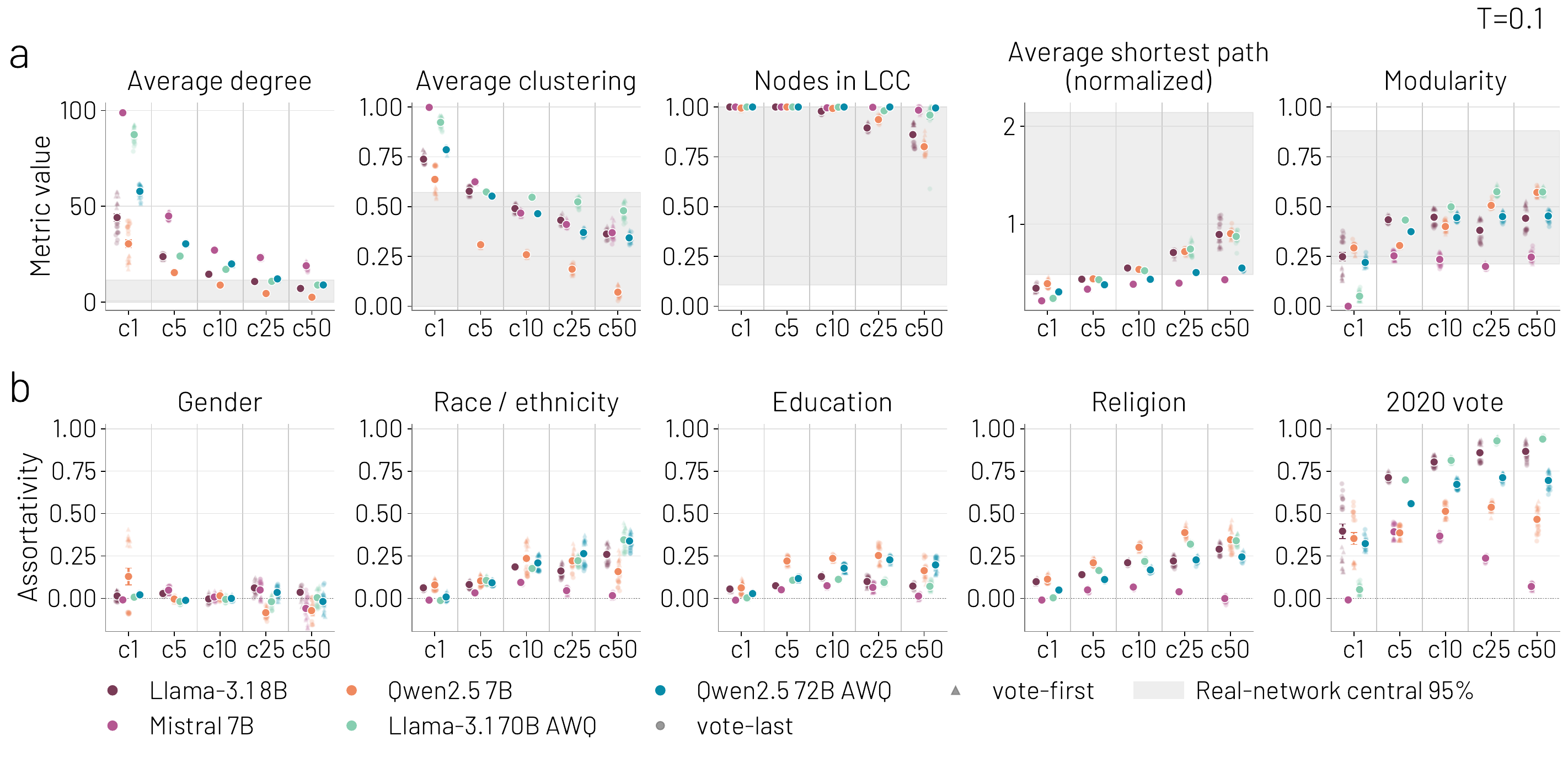}
\caption{\textbf{Network metrics vary across models and prompting techniques.} Prompt versions contain an increasing number of candidate friends, from one (c1) to fifty (c50), and differ in the ordering of persona attributes, mentioning the presidential vote first or last (circle vs. triangle markers). Larger colored points show model averages across attribute orders and runs ($10$ per configuration), and error bars indicate standard errors. Results are shown for temperature $T=0.1$. The gray area represents the plausible values of the metrics according to the empirical benchmarks. \textbf{(a)} Structural network metrics vary most under the pairwise prompt (c1), both across models and, for some models, across runs and attribute orderings. Among the list-based prompts, larger candidate sets produce sparser networks with lower clustering and longer shortest paths, while networks remain almost entirely connected. Modularity also tends to increase with larger candidate sets, although this pattern varies across models. \textbf{(b)} Assortativity is limited for demographic attributes, although increasing with the number of candidates in the prompt, whereas 2020 presidential vote shows stronger assortative mixing and larger differences across models and prompting configurations.}
\label{fig2}
\end{figure}

To assess how the structure of the generated networks varies and whether these networks reproduce structural properties observed in empirical social networks, we compare reciprocal networks (see Methods section) with those observed across 57 friendship and friendship-like networks from 13 empirical datasets, following previous work~\cite{chang_llms_2025}.
Here, reciprocal networks include a tie between two personas only when both nominate each other as potential friends.
To this end, we consider five main network metrics: (i) average degree, which measures the average number of ties per node, (ii) the average clustering coefficient, which captures the tendency toward local clustering, (iii) the proportion of nodes in the largest connected component (LCC), (iv) the average shortest path length, and (v) Louvain modularity, which measures the extent to which the network separates into communities.
A generated network is considered structurally plausible when all five measures fall within the corresponding empirical ranges (see Methods).
We also compute assortativity by demographic and political attributes to assess whether generated ties preferentially connect individuals with similar characteristics.
Fig.~\ref{fig2} reports these metrics across models and prompting configurations at a low model temperature of $T=0.1$.
The analysis with $T=0.7$, nomination networks, and complete results with standard errors can be found in the SI (Figs.~S1-S2 and Tables~S2-S5).

Figure~\ref{fig2}a shows how network structure differs across prompting configurations, and in particular based on the number of potential friends included in the prompt.
The pairwise configuration (c1) shows the largest variability between models and, for some models, between runs and attribute orderings, generating networks whose structural properties vary even within the same model.
Mistral 7B tends to produce nearly complete networks, with density close to 1, whereas Llama-3.1 8B and Qwen2.5 7B show the highest variation across runs.
As a result, c1 also yields a wide variation in the other structural metrics, including clustering, shortest-path length, and modularity. 
In contrast, the list-based methods (from c5 with 5 candidates to c50 with 50 candidates) produce more stable network structures across runs and configurations.
As the number of candidates in the prompt increases from c5 to c50, average degree and clustering decrease, while the normalized average shortest path length and modularity generally increase.
Networks remain largely connected across configurations, although the proportion of nodes in the LCC decreases for some models with larger sets of candidates.
On average, these changes move the structural properties closer to the empirical ranges, especially for c25 and c50, meaning that the number of candidates in the prompt strongly affects the structure of the generated networks.

Based on these measures, 332 of the 1,000 reciprocal networks ($33.2\%$) fall within the empirical ranges for all five metrics.
The proportion of plausible networks increases from $2\%$ for c1 and $0\%$ for c5 to $20\%$ for c10, $64\%$ for c25, and $80\%$ for c50.
In this evaluation, the average degree is the most restrictive criterion, with all generated networks whose average degree falls within the empirical range also falling within the ranges for the other four measures.
Differences also emerge across models, with Qwen2.5 7B having the highest proportion of plausible networks ($61\%$), followed by Llama-3.1 8B ($40.5\%$), Llama-3.1 70B ($40\%$), and Qwen2.5 72B ($24.5\%$), with none of the Mistral 7B networks meeting the criteria.

Figure~\ref{fig2}b shows the results on attribute assortativity.
Gender assortativity remains close to $0$ in most models and list-based configurations, while race and ethnicity, education, and religion show increasing assortativity as the number of candidates in the prompt increases and the average degree decreases.
By contrast, and consistent with previous work~\cite{chang_llms_2025,kilaru_when_2026}, assortativity by political vote is considerably higher, though it varies widely across models and prompting configurations.
Among the list-based methods, the two Llama models show high political assortativity, which increases with the number of candidates, while Qwen2.5 7B and 72B have lower but still high values.
Mistral 7B instead shows lower political assortativity across most configurations, with values decreasing for larger candidate sets.
Some models, such as Llama-3.1 8B and Qwen2.5 7B and 72B, are also sensitive to attribute order, showing higher vote assortativity when vote is presented first in the persona description (see triangle markers in Fig.~\ref{fig2}b).

\subsection{Survey-based evaluation reveals model-specific errors in partisan exposure}

While political assortativity captures the tendency of generated ties to connect politically similar individuals, the measure does not indicate whether these patterns reproduce the partisan environments reported by individuals.
We therefore move from network-level sorting to ego-level exposure using the nomination networks (see Methods), which include all outgoing friendship choices made by each persona without requiring that the friendship choice is reciprocated by the alter.
This permits us to capture directional differences in friendship choices, including cases in which an ego nominates an alter who does not reciprocate the choice, which may be lost when considering only reciprocal ties.
Specifically, we compare exposure to Democrats and Republicans in the generated networks with the corresponding ANES measures in the survey, as described in the Methods section.
For each ego, we compute the proportions of ties to Republican and Democratic alters and bin these proportions into 5 categories to match the 1--5 Likert scale of the survey question ``How many of your friends or family are Democrats/Republicans?''.

\begin{figure}[t!]
\centering
\includegraphics[width=1\textwidth]{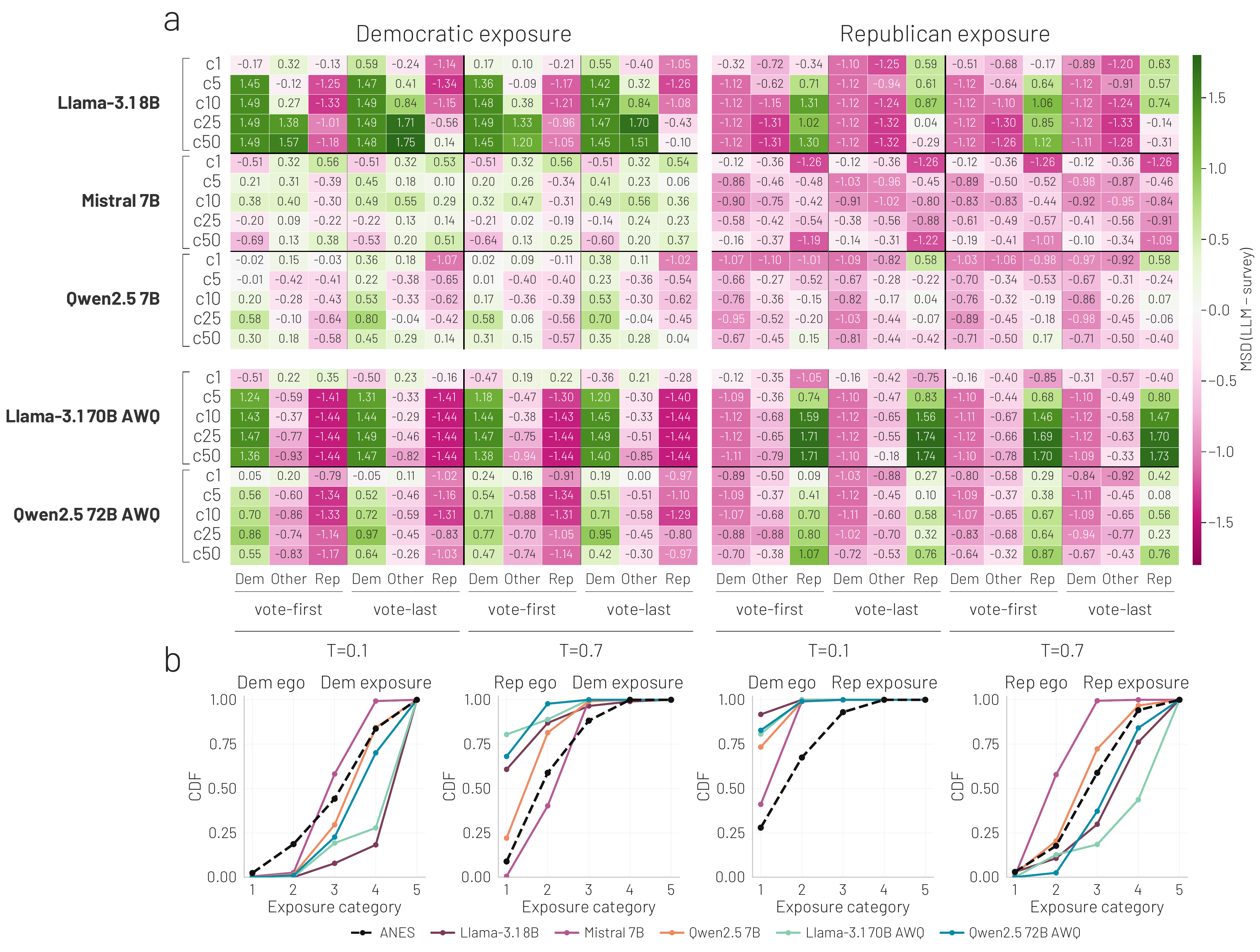}
\caption{\textbf{Distance between partisan exposure in LLM-generated networks and ANES survey using the nomination networks.} \textbf{(a)} Mean Signed Deviation (MSD) between partisan exposure in the LLM-generated networks and in the ANES survey across models and experimental configurations. Mistral 7B and Qwen2.5 7B tend to underestimate Republican exposure, while the two Llama models and Qwen2.5 72B to overestimate same-party exposure and underestimate cross-party exposure; \textbf{(b)} Cumulative distribution functions of the average Democratic and Republican exposure across models compared to the survey measure. Here, lines below the survey one suggest an overestimation of the exposure, while lines above the survey measure represent an underestimation.}\label{fig3}
\end{figure}

Figure~\ref{fig3} shows the comparison between LLM-generated and survey-reported partisan exposure.
Panel \textit{a} reports the Mean Signed Deviation (MSD) between generated and survey-reported exposure across models and experimental configurations.
Negative values indicate an underestimation relative to the survey, while positive values indicate an overestimation, in terms of points on the 1--5 exposure scale.
This is summarized by panel \textit{b}, which shows the corresponding cumulative distributions for Democratic and Republican egos' exposure to Democratic and Republican alters.

Here, we consider all generated nomination networks, rather than restricting the analysis to runs whose reciprocal networks are structurally plausible, because the aim is to evaluate how the friendship choices generated by different models and configurations translate into egos' partisan exposure.

The comparison with the survey shows that generated social networks differ from real data in both the direction of the error and the type of partisan environment they generate.
Some models generate networks with a stronger separation between Democratic and Republican egos than reported in the survey, with higher same-party and lower cross-party exposure. Other models underestimate Republican exposure across ego groups.

Specifically, the two Llama models and Qwen2.5 72B generate a stronger separation between Democratic and Republican egos than reported in the survey.
The largest deviations are observed for Llama-3.1 70B, where differences across ego--alters relations exceed one point on the 1--5 exposure scale in both directions, and in Llama-3.1 8B, suggesting that this stronger separation is not limited to the larger models.
Also Qwen2.5 72B shares the same general pattern, but with smaller deviations.

On the other hand, Mistral 7B and Qwen2.5 7B show a different pattern, characterized mainly by an underestimation of Republican exposure across ego groups and most configurations. 
For both models, exposure to Republican alters is generally lower than reported in the survey, including for Republican egos, while deviations in Democratic exposure are smaller and more mixed.

The pairwise configuration (c1), whose reciprocal networks are rarely structurally plausible, represents the main exception to these model-specific patterns. 
Specifically, for several models, c1 produces exposure values that differ from the list-based configurations and are closer to the survey-reported measures.

Differences across experimental configurations are on average smaller than those observed across models.
Among the list-based configurations, list size (i.e., the number of alters in the prompt) changes the magnitude of the deviations in some cases.
However, the main direction of the results remains relatively stable across chunk sizes.
Similarly, changing the order in which demographic and political attributes are presented produces limited differences in the overall pattern.
Finally, results are also consistent across temperatures ($0.1$ and $0.7$).

To understand how reciprocity changes the partisan environments generated around personas, we repeat the analysis using reciprocal networks (SI, Fig. S3 and Table S9).
Compared with the nomination networks, reciprocal ties produce a clearer separation between Democratic and Republican egos across all models, with Democratic egos more exposed to Democratic alters and Republican egos more exposed to Republican alters.
This suggests that considering only mutually selected ties can change the egos' partisan environments, while hiding some of the asymmetries and differences that characterize the friendship choices of each model.

\subsection{Political assortativity does not fully capture ego-level partisan exposure}
The previous analysis shows that different models and configurations can generate different partisan environments at the individual level when compared with the ANES survey.
We next examine whether networks with similar political assortativity, which summarizes political mixing at the network level, also generate similar partisan environments for the same personas.

\begin{figure}[t]
\centering
\includegraphics[width=1\textwidth]{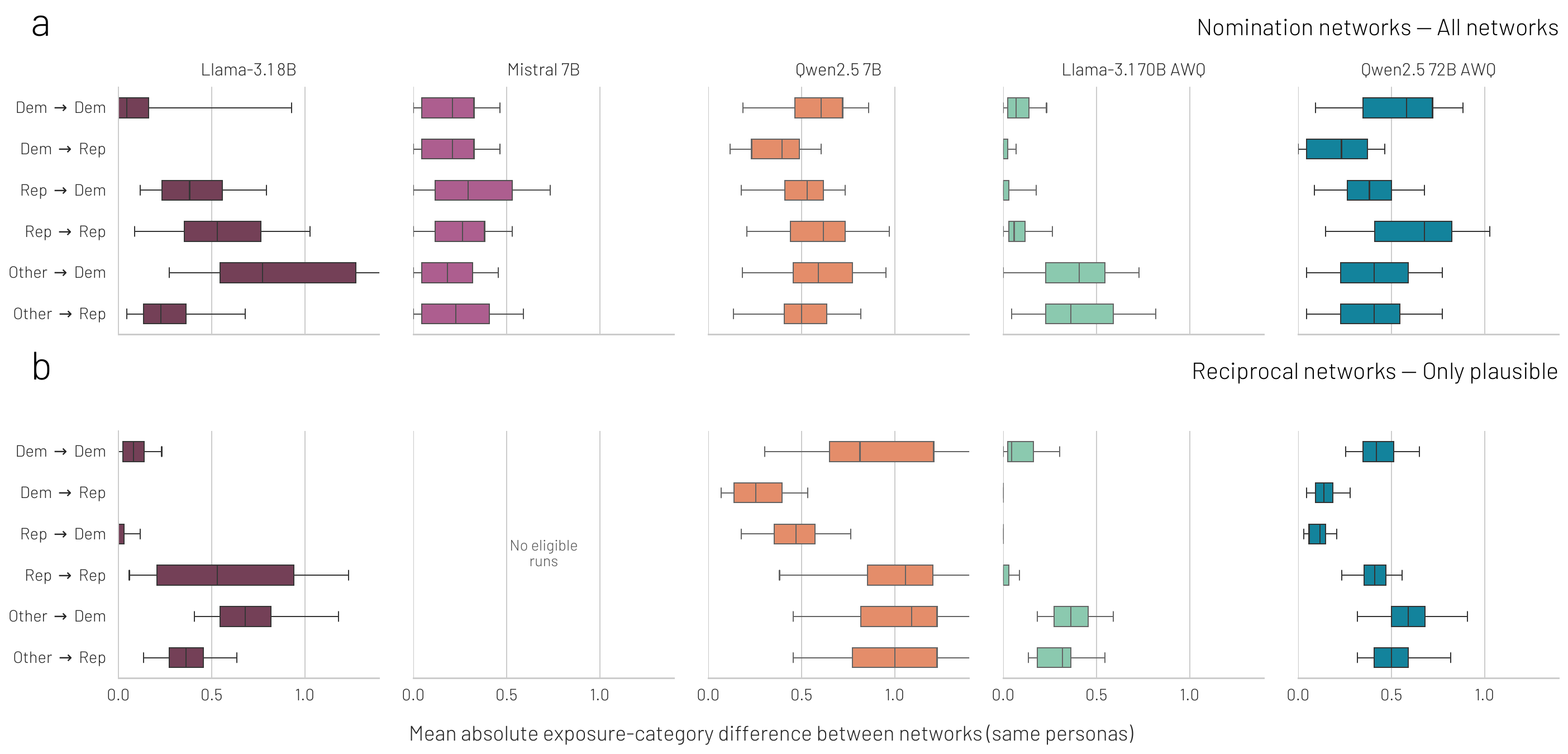}
\caption{\textbf{Exposure differences among networks with similar political assortativity.} Mean absolute within-person differences in partisan exposure between pairs of LLM-generated networks from the same model whose vote assortativity differs by less than \(0.02\). For each pair, exposure is compared for the same personas on the 1--5 exposure scale. Results are shown using the nomination networks \textbf{(a)} and the reciprocal networks that meet all five plausibility criteria \textbf{(b)}.
Boxes show the interquartile range, center lines indicate medians, and whiskers the 5th--95th percentiles. Networks can differ in prompting configuration.}\label{fig4}
\end{figure}

To this end, within each model, we compare pairs of networks generated under different experimental configurations whose political assortativity differs by less than $0.02$.
The number of matched network pairs for each model and threshold is reported in SI, Table S10.
For each pair, we compare the exposure categories generated for the same personas and compute the mean absolute difference, separately for Democratic, Republican, and Other egos' exposure to Democratic and Republican alters (Figure~\ref{fig4}).
Here, higher values indicate larger differences in partisan exposure around the same individuals despite similar levels of network-level political assortativity.
The results are consistent with a threshold equal to $0.05$ (see SI, Fig. S4).

We perform this comparison using both network representations.
For nomination networks, representing each ego's friendship choices, we consider all generated networks (Figure~\ref{fig4}a), consistent with the exposure analysis in the previous analysis.
On the other hand, for reciprocal networks, for which structural plausibility can be evaluated against the empirical friendship-networks benchmarks, we consider only the networks that meet the five plausibility criteria (Figure~\ref{fig4}b), as described in previous analyses.
This allows us to assess whether differences in ego-level exposure among networks with similar political assortativity are limited to implausible networks or extend also to networks with plausible structural properties.

In the nomination networks, the results show that networks with similar levels of political assortativity can still generate different partisan exposure for the same personas, although the magnitude of these differences varies across models and ego--alter combinations (Fig.~\ref{fig4}a).
Llama-3.1 70B shows relatively small differences for Democratic and Republican egos, with larger differences for Other egos, while Llama-3.1 8B also shows larger differences for Republican and Other egos. 
Mistral 7B shows more homogeneous intermediate differences across relations, whereas the two Qwen models show larger differences for several ego--alter combinations.
Even though many of these values are below one, they represent differences measured on a 1--5 exposure scale, corresponding to meaningful changes in the partisan environments generated around the same personas, reaching differences of one full point of the scale.

Importantly, these differences do not appear to be due to structurally implausible networks.
Among reciprocal networks that meet all five criteria (Fig.~\ref{fig4}b), ego-level exposure still varies between networks with similar political assortativity.
The differences remain relatively small for Democratic and Republican egos in Llama-3.1 70B, but are larger for specific relations in Llama-3.1 8B and Qwen2.5 72B, and are particularly evident for Qwen2.5 7B, where several ego--alter relations differ by around one point on average.
Here, Mistral 7B is not included because none of its reciprocal networks satisfies all five plausibility criteria. 
Therefore, the observed differences between aggregate political assortativity and ego-level partisan exposure cannot be attributed only to networks with implausible structural properties.

\subsection{Group-specific errors in LLM-generated partisan exposure}\label{results:biases}
\begin{figure}[t]
\centering
\includegraphics[width=1\textwidth]{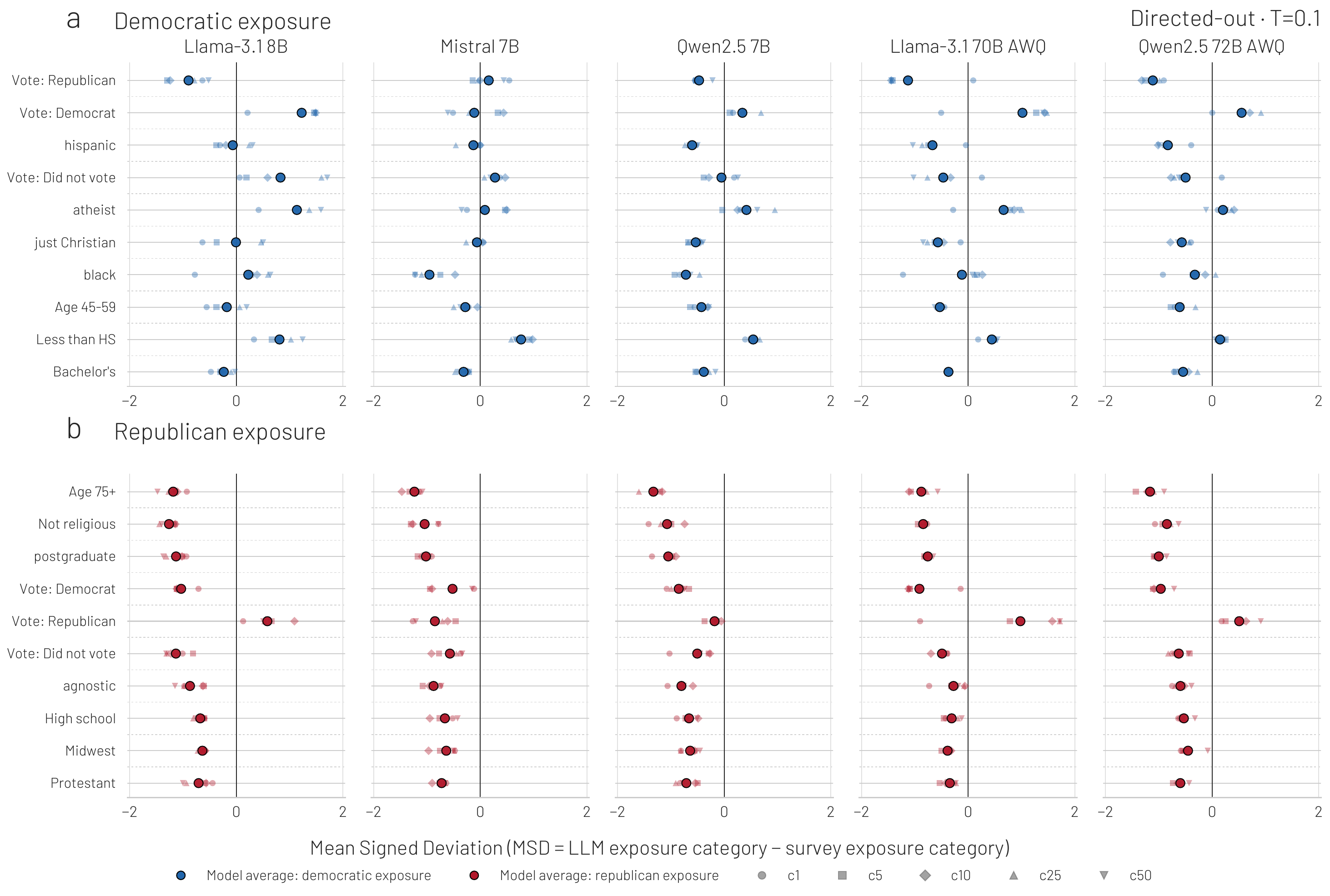}
\caption{\textbf{Group-specific errors in partisan exposure in LLM-generated networks.} Mean Signed Deviation (MSD) between LLM-generated and survey-reported exposure to \textbf{(a)} Democrats and \textbf{(b)} Republicans for the social groups with the largest deviations. Democratic exposure shows group-specific patterns of under- and overestimation, while Republican exposure is mainly underestimated. Different point styles represent the different prompting methods, with larger markers showing the model average. Results are shown for $T=0.1$. }\label{fig5}
\end{figure}

We also assess whether errors in the partisan exposure of LLM-generated networks are associated with specific social groups.
To this end, we compute the Mean Signed Deviation (MSD) between generated networks and survey-reported exposure for demographic and political groups. 
Figure~\ref{fig5} shows the demographic and political groups with the largest deviations for Democratic and Republican exposure, with negative values indicating an underestimation of partisan exposure relative to the survey, while positive values indicate an overestimation.
Results are shown for $T=0.1$, with results for $T=0.7$ reported in the SI (Fig.~S5).
Complete group-specific results are reported in SI, Table S11.

For Democratic exposure (Fig. \ref{fig5}a), both the direction and magnitude of the errors vary across demographic and political groups and models.
Exposure to Democrats is underestimated for Hispanic respondents, respondents aged 45--59, and those with at least a bachelor's degree, across all the models.
Instead, Democratic exposure for African Americans is underestimated in all the models considered, except for Llama-3.1 8B.
In contrast, Democratic exposure is overestimated among respondents with lower educational attainment (i.e., less than high school), and atheists.
Substantial differences also emerge across political groups, with the two Llama models and Qwen2.5 72B overestimating Democratic exposure among Democratic voters and underestimating it among Republican voters.
Interestingly, these differences do not disappear with increasing model size, but remain relatively consistent across models.

As suggested by the previous analysis, the pattern is more directional for the exposure to Republicans (Fig. \ref{fig5}b), with most of the selected groups that show negative deviations from the survey.
Exposure to Republicans is consistently underestimated among respondents aged 75 or older, respondents with postgraduate and high school education, Protestants, agnostics, and non-religious respondents, and those living in the Midwest.
Republican exposure is also underestimated among Democratic voters and respondents who did not vote, while Republican voters represent the main exception, with the two Llama models and Qwen2.5 72B that overestimate their exposure to Republicans, while Mistral 7B and Qwen2.5 7B underestimate it.
These patterns are visible in both smaller and larger models and across configurations, although their magnitude varies.

Overall, these results show that errors in partisan exposure vary across demographic and political groups, with group-specific patterns of underestimation and overestimation that persist across models and configurations.

\section{Discussion}\label{discussion}
We evaluated LLM-generated political networks through their structural properties, political assortativity, and the comparison between ego-level partisan exposure and survey-reported exposure.
Our findings show that, even if LLMs can generate networks with plausible structural properties, these properties strongly depend on the model, parameters, and prompting techniques used for the tie formation, and do not follow a clear relationship with model size.
When comparing generated partisan exposure with the ANES survey, we find that the partisan environments generated for individual personas can differ from those reported by respondents.
Specifically, some models generate stronger separation between Democratic and Republican egos than reported in the ANES survey, with higher same-party and lower cross-party exposure, while others mainly underestimate Republican exposure across ego groups.
In this context, increasing model size does not necessarily lead to generated partisan environments that are closer to the survey benchmark, with the larger models considered that generate stronger separation between Democratic and Republican egos.
This is consistent with recent work showing that model scale can change the structural and relational patterns generated by LLMs without implying improvements in realism~\cite{kilaru_when_2026}.
At the same time, our results show that similar levels of political assortativity do not necessarily imply that the partisan environments experienced by individuals are also reproduced correctly.
This is because while political assortativity summarizes the overall tendency of ties to connect politically similar individuals, it does not indicate how exposure is represented across Democratic, Republican, and other egos and alters.
As a consequence, similar levels of assortativity can correspond to different patterns of same-party and cross-party exposure, also among plausible reciprocal networks.
Moreover, these errors are not distributed uniformly across the population, as specific demographic and political groups show consistent patterns of under- and overestimation in partisan exposure across models and configurations, extending previous concerns about the representation of demographic personas in LLMs \cite{santurkar_whose_2023,cheng_compost_2023,lee_large_2024,wang_large_2025,coppolillo2026unmasking}.
In this context, errors in individual tie formation can therefore accumulate into group-specific differences in partisan exposure, even when average network properties appear plausible.
Therefore, aggregate measures can hide representational errors in who is exposed to whom, in which direction, and for which social groups, even when the generated network appears plausible from a structural perspective.

These findings connect to concerns about the empirical validity of LLM-based social simulations, with previous studies that have shown that LLMs may reproduce aggregate distributions or collective outcomes while having different patterns at the individual or subgroup levels.
For example, LLM-based populations can reproduce collective cooperation dynamics, while failing to capture individual heterogeneity and conditional behavior~\cite{arruda_collective_2026}, while synthetic survey responses can approximate overall population averages, but with lower variation and distorted patterns across demographic groups~\cite{bisbee_synthetic_2024}.
Moreover, this micro--macro mismatch has also been shown in multi-agent settings, where interactions among LLM agents can generate emergent collective patterns and population-level biases that differ from those observed at the level of individual agents~\cite{ashery_emergent_2025,flint_group_2026}.
Although these studies focus on repeated interactions among agents, our results show that the difference between micro and macro outcomes can also emerge in a simpler tie formation setting, where plausible aggregate network properties can hide errors in the partisan environments generated at the ego level.
This also supports recent calls for empirical benchmarking of LLM-based social simulations~\cite{munker_dont_2026}.

This distinction is also relevant for political networks, where the composition of individuals' political environments and the political context can be associated with their political behavior~\cite{huckfeldt_citizens_1995,sinclair_social_2012,huckfeldt_networks_2014,butters_polarized_2022,tonin_physical_2025}. 
Our results show that some models misrepresent these political environments. Some overestimate same-party exposure and underestimate cross-party exposure, generating stronger partisan segregation. Other models mainly underestimate exposure to Republicans, producing asymmetric errors.
Therefore, errors in micro-level tie formation decisions can alter the political contexts generated around personas, 
and may affect simulated social and political processes such as influence, persuasion, communication, or polarization.

The structural properties of the generated networks also depend on how the tie formation task is implemented, with pairwise and list-based prompts producing different network structures.
This is consistent with recent work showing that prompting architecture and methodological choices can affect the structural and assortative properties of LLM-generated networks~\cite{chang_llms_2025,kilaru_when_2026}.
Therefore, the ability of LLMs to generate realistic networks strongly depends on how the tie formation task is operationalized and should not be interpreted as a property of the models.

Our work comes with limitations.
First, the survey measures from the ANES 2020-2022 Social Media Study are self-reported answers of respondents' exposure to Democratic and Republican friends or family and do not provide direct observations of complete social networks. 
However, they provide an empirical measure of respondents' perceived partisan environments that can be used to compare the partisan exposure generated for the corresponding survey-based personas.
Second, due to the computational cost of generating all possible ties across multiple models, runs, temperatures, prompting methods, and attribute orderings, we use a reduced sample of ANES respondents (i.e., $N=100$).
Although the sampling procedure preserves the main demographic and political composition of the survey, this step could reduce heterogeneity in the sample.
Third, the personas include a limited set of demographic and political attributes, while friendship formation depends on a wider range of individual characteristics, social contexts, and opportunities for interaction~\cite{mcpherson_birds_2001,kossinets_origins_2009,rivera_dynamics_2010,smith_social_2014}.
Fourth, our analysis considers a limited set of open-weight instruction-tuned models that can be deployed locally, including quantized versions of the larger models, which can affect research results in ways that are difficult to predict~\cite{ferrara_quantibias_2026}.
Therefore, the observed differences across model families and sizes may not generalize to proprietary models, base models, or other instruction-tuned architectures.
Finally, the tie formation task represents friendship as a one-step decision, along with the undirected reciprocal networks, and does not capture the interactive and dynamic processes through which social relationships emerge, persist, and change over time.

Despite these limitations, our findings show that plausible aggregate outcomes in LLM-based social simulations can hide errors at the individual level.
In our setting, plausible social network properties and similar levels of political assortativity can hide differences in the partisan environments generated around individuals and across social groups.
This macro--micro mismatch highlights the importance of evaluating LLM-based social simulations at both the aggregate and individual levels, especially when micro-level decisions and interactions contribute to the collective outcomes being studied.

\section{Materials and Methods}\label{methods}

\subsection{Data sources and sampling}
We use data from the American National Election Studies (ANES) 2020-2022 Social Media Study \cite{anes_socialmedia_2023}, which includes 5,750 respondents in the 2020 pre-election wave and 5,277 in the 2020 post-election wave. 
Our analysis focuses on respondents with valid post-stratification weights and available information on demographic characteristics, political preferences, and self-reported social exposure.
To characterize the personas used in the tie formation tasks we consider age, gender, race and ethnicity, education, region, religion, and vote choice.
The survey also captures the composition of the respondents' social environments, including the number of friends or family members identified as Democrats or Republicans on a Likert scale from 1 (none of them) to 5 (all of them), which we use as empirical benchmark for ego-level partisan exposure.
From the survey, we keep respondents with valid post-stratification weights, valid survey-based partisan exposure measures, and complete information for the demographic and political attributes used to construct the personas.
Due to the computational costs of generating ties across all possible pairs of respondents and repeating the procedure across multiple models and prompt configurations, we construct a reduced sample of 100 respondents.
To this end, we use weighted stratified sampling based on gender, age group, race and ethnicity, and education. 
For each joint stratum, we compute its share of the eligible ANES population using the survey post-stratification weights and allocate the target sample proportionally to this weighted share.
Respondents are then sampled within each stratum with probability proportional to their post-stratification weight.
We additionally compare the resulting sample with the weighted ANES population across the stratification variables, region, religion, and 2020 presidential vote.
The distributions remain close to the weighted survey population (SI, Table~S1).

\subsection{Models}
To perform the tie formation tasks we use open-weight instruction-tuned LLMs that can be run locally on a single NVIDIA A40 GPU.
Specifically, we use five models: Llama-3.1-8B-Instruct, Qwen2.5-7B-Instruct, Mistral-7B-Instruct-v0.3, Llama-3.1-70B-Instruct-AWQ4, and Qwen2.5-72B-Instruct-AWQ4.
For each model, we generate ties using two temperature values, $0.1$ and $0.7$.

This local setup is important given the computational cost of the task.
Generating a network requires evaluating potential ties among all sampled personas, and the procedure is repeated across models, temperatures, prompt formats, chunk sizes, and attribute orderings.
For this reason, we use models that can be deployed on local GPU resources, including quantized versions of larger models when needed, such as Llama-3.1-70B-Instruct-AWQ4, and Qwen2.5-72B-Instruct-AWQ4.

\subsection{Persona construction and tie formation task}
For each sampled respondent, we construct a persona using age, gender, race and ethnicity, education, state of residence, religion, and vote choice in the 2020 U.S. presidential election (see Figure~\ref{fig1}a).
We use two attribute orderings to test whether the position of political information in the persona description affects tie formation (i.e., vote as first or last attribute in the description).
We then ask the model to decide whether the persona would consider another persona, or list of personas, as friends. 
We use two prompting formats, as shown in Fig. \ref{fig1}b. 
In the pairwise configuration (c1), the model evaluates one potential alter at a time and returns a binary friendship decision. 
In the list-based format, the model evaluates multiple potential alters within the same prompt. 
For the list-based task, we vary the number of candidates included in each prompt using sets of 5, 10, 25, and 50 candidates (c5, c10, c25, and c50). 
For each candidate set, the prompt presents the ego persona followed by the IDs and attributes of the candidate alters, and asks the model to return only the IDs of the individuals it would befriend. 
Each potential alter is therefore evaluated once for each ego in every generated network, while self-ties are excluded.

For each generated run, we construct two network representations from these friendship decisions.
First, we construct a nomination network, where a directed edge from the ego $i$ to alter $j$ is created when the model returns a positive friendship decision. 
This network therefore includes all ego-to-alter nominations, representing the set of potential friends selected by each persona.
However, because each ego makes friendship choices independently, the model might select alters who would not reciprocate the choice, producing unreciprocated nominations that reflect an artifact of the friendship decision~\cite{kitts_what_2021}.
Second, we construct a reciprocal network, where an undirected tie between two personas is created only when both nominate each other, capturing mutually selected friendships and providing a more direct comparison with empirical friendship networks.
In the study, we primarily use the nomination networks to measure ego-level partisan exposure, while reciprocal networks are used for the structural evaluation.

Each configuration is repeated across $10$ runs, changing the random seed to account for variation in the model outputs.
This results in $1,000$ generated runs across the five models, five prompting methods, two temperatures, two attribute orderings, and ten repetitions.

\subsection{Network measures and structural plausibility}
We evaluate the structural properties of the generated networks using reciprocal friendship ties, creating an undirected tie between two personas when both nominate each other, which allows us to compare the generated networks with empirical friendship networks.
We characterize the generated networks using five structural measures: average degree, average clustering coefficient, the proportion of nodes in the largest connected component (LCC), average shortest path length, and Louvain modularity.

The average clustering coefficient is computed across all nodes. 
The average shortest-path length is computed on the largest connected component, normalized by its size. 
Finally, we identify communities using the Louvain algorithm and compute the corresponding modularity.

To evaluate the structural plausibility of the generated networks, we compare these measures with a reference set of $57$ friendship and friendship-like networks from $13$ empirical datasets.
These include Baerveldt Dutch Social Behavior~\cite{houtzager_just_1999}, Galesburg~\cite{coleman_diffusion_1957}, Hi-tech~\cite{krackhardt1999}, High School 2013~\cite{mastrandrea_contact_2015}, Kapferer Tailor Shop~\cite{kapferer_strategy_1972}, Knecht class 12b~\cite{knecht_friendship_2010}, Lazega Lawyers~\cite{lazega_collegial_2001}, MSW Students~\cite{mauldin_relationships_2017}, Moreno High School~\cite{rossi_network_2015}, PEAS~\cite{oosterhoff_early_2022}, Prison~\cite{macrae_direct_1960}, Sciences Po~\cite{algan_friendship_2026}, and Zachary Karate~\cite{zachary_information_1977} (see SI, Table S6).
We use these empirical networks to define reference ranges for each structural measure.
When a dataset includes multiple networks, we assign equal total weight to each dataset and divide this weight across the networks that belong to it. 
We then use the weighted $2.5th$ and $97.5th$ percentiles of each structural measure to define the empirical reference ranges.
The resulting empirical ranges are reported in SI, Table S7.
A generated network is classified as structurally plausible when all five measures fall within the corresponding ranges.
Density is not included in this classification because it depends directly on network size.

We also measure assortative mixing by gender, race and ethnicity, education, religion, and vote choice in the 2020 U.S. presidential election by computing the attribute assortativity for each attribute.
Positive values indicate that connected personas tend to share the same attribute more frequently than expected from the corresponding mixing pattern, while values around zero indicate limited assortative mixing.

\subsection{Validation of the partisan exposure using the ANES survey}
To evaluate how the LLM-generated networks capture individual-level partisan exposure, we compared ego-level metrics from the simulated networks with responses to the ANES 2020-2022 Social Media Study survey question: ``How many of your friends or family are Democrats/Republicans?'', which uses a Likert scale ranging from $1$ (\textit{None of them}) to $5$ (\textit{All of them}).

Our primary measure of generated exposure uses the friendship nominations made by each ego.
For each ego, we compute the proportion of outgoing ties directed toward Democratic and Republican alters.
The denominator includes all outgoing alters with a valid political affiliation, including those classified as neither Democratic nor Republican.
For egos with no valid outgoing ties, exposure is set to zero.
To compare the generated exposure with the survey, we transform these proportions into five equally spaced categories.
Exposure between 0 and 0.2 is assigned to category 1, between 0.2 and 0.4 to category 2, between 0.4 and 0.6 to category 3, between 0.6 and 0.8 to category 4, and between 0.8 and 1 to category 5.
We then compare, for each ego, the generated exposure category with the corresponding survey response.
Additional information on partisan exposure can be found in the cumulative distribution functions in Figure~\ref{fig3}b and in the SI, Table~S8.
For each ego $i$ and exposure to party $p$, we define exposure error as follows.

\begin{equation}
e_{i,p} = E^{LLM}_{i,p}-E^{survey}_{i,p},
\end{equation}

where $E^{LLM}_{i,p}$ is the generated exposure category for ego $i$ to party $p$, and $E^{survey}_{i,p}$ is the corresponding ANES response.
We then quantify the difference between partisan exposure in the LLM-generated networks and in the survey by computing the Mean Signed Deviation (MSD), defined as the average difference between generated and survey-reported partisan exposure, as follows.

\begin{equation}
MSD_p = \frac{1}{N}\sum_{i=1}^{N}
\left(E^{LLM}_{i,p}-E^{survey}_{i,p}\right),
\end{equation}

Positive values indicate that the generated networks overestimate exposure to Democrats or Republicans, while negative values indicate that they underestimate exposure.
We compute MSD separately for Democratic, Republican, and Other egos and for exposure to Democratic and Republican alters.
Finally, we compare the distributions of generated and survey-reported exposure using cumulative distribution functions, and we repeat the analysis using reciprocal networks.

\subsection{Political assortativity and ego-level partisan exposure}
To assess whether networks with similar levels of political assortativity also generate similar partisan environments around individuals, we compare pairs of generated networks whose vote assortativity differs by less than $0.02$, and we repeat the analysis using a threshold of $0.05$ in the SI.
We perform the matching separately within each model and compare networks generated under different experimental configurations.
For nomination networks, we consider all generated networks, while for reciprocal networks we restrict the analysis to those satisfying all five structural plausibility criteria described above. 
In both cases, vote assortativity and partisan exposure are computed on the same network representation.

For each pair of networks $A$ and $B$, we compare partisan exposure for the same personas that have valid exposure measures in both networks.
Specifically, for ego $i$ and exposure to party $p$, we compute the absolute difference between the exposure categories in the two networks and average these differences across the personas included in the networks:

\begin{equation}
D_{AB,p} =
\frac{1}{N_{AB}}
\sum_{i=1}^{N_{AB}}
\left|
E^{LLM}_{iA,p} - E^{LLM}_{iB,p}
\right|,
\end{equation}

where $N_{AB}$ is the number of personas in the corresponding ego group with valid exposure measures in both networks.
Higher values indicate larger differences in partisan environments generated around the same individuals, despite similar political assortativity.

We report the distribution of this difference for Democratic, Republican, and Other egos' exposure to Democratic and Republican alters.
Since a generated network can be included in multiple pairs satisfying the assortativity threshold, these comparisons are used descriptively rather than as independent observations.

\subsection{Group-specific errors in partisan exposure}
To assess whether errors in partisan exposure vary across social groups, we compute the Mean Signed Deviation separately for respondents with different demographic and political characteristics using the nomination networks.
We consider gender, age, race and ethnicity, education, region, religion, and vote choice in the 2020 U.S. presidential election.
We exclude groups represented by fewer than five personas in the sampled population.

For each group $g$ and exposure to party $p$, we compute

\begin{equation}
MSD_{g,p} = \frac{1}{N_g}\sum_{i \in g}
\left(E^{LLM}_{i,p}-E^{survey}_{i,p}\right),
\end{equation}

where $N_g$ is the number of respondents belonging to group $g$.
Positive values indicate an overestimation of exposure to the corresponding party for that group, while negative values indicate an underestimation relative to the survey.
We first compute the MSD separately for each generated network and social group and then average these values across runs.

\section*{Code and data availability}
The American National Election Studies (ANES) 2020-2022 Social Media dataset is available in the American National Election Studies website~\cite{anes_socialmedia_2023}.
Data and code to reproduce the main results of the study are publicly available at \url{https://github.com/tonmarco/llm-social-networks-replication}.

\section*{Acknowledgements}
M.To. acknowledges the support of the NRRP MUR program funded by the NextGenerationEU. 
The work of B.L. was partially supported by the following projects: Horizon Europe Programme, grant \#101120237-ELIAS and grant \#101120763-TANGO. Funded by the European Union. L.M.A. acknowledges the support of the Carlsberg Foundation through the COCOONS project under grant number CF21-0432. Views and opinions expressed are however those of the author(s) only and do not necessarily reflect those of the European Union or the European Health and Digital Executive Agency (HaDEA). Neither the European Union nor the granting authority can be held responsible for them.

\section*{Author contributions}
M.To. conducted the experiments and drafted the manuscript.
M.To. and L.M.A. designed the experiments.
M.To., M.Ti., B.L., and L.M.A. designed the study, interpreted the results, provided critical feedback, and revised the manuscript.
All authors approved the final version of the manuscript.

\section*{Competing interests}
The authors declare no competing interests.





\bibliography{sn-bibliography}

\end{document}


\title{Supplementary Information for: Validating LLM-generated political networks with survey-reported partisan exposure}


\author*[1,2]{\fnm{Marco} \sur{Tonin}}

\author[1,2]{\fnm{Michele} \sur{Tizzoni}}

\author[2]{\fnm{Bruno} \sur{Lepri}}

\author[3]{\fnm{Luca Maria} \sur{Aiello}}

\affil[1]{\orgdiv{Department of Sociology and Social Research}, \orgname{University of Trento}, \orgaddress{\street{Via Verdi 26}, \city{Trento}, \postcode{I-38122}, \country{Italy}}}

\affil[2]{\orgdiv{Mobile and Social Computing Lab (MobS)}, \orgname{Fondazione Bruno Kessler}, \orgaddress{\street{Via Sommarive 18}, \city{Trento}, \postcode{I-38123}, \country{Italy}}}

\affil[3]{\orgdiv{Data Science Section}, \orgname{IT University of Copenhagen}, \orgaddress{\street{Rued Langgaards Vej 7}, \city{Copenhagen}, \postcode{DK-2300}, \country{Denmark}}}


\maketitle

\renewcommand{\thesection}{S\arabic{section}}
\renewcommand{\thesubsection}{S\arabic{section}.\arabic{subsection}}

\renewcommand{\thefigure}{S\arabic{figure}}
\renewcommand{\thetable}{S\arabic{table}}
\renewcommand{\theequation}{S\arabic{equation}}

\setcounter{tocdepth}{2} 
\tableofcontents

\newpage
\section{Sample}
\label{sec:sample}
Table \ref{tab:sample_composition} compares the demographic and political composition of the $N=100$ sample used to generate the networks with the weighted distribution of the eligible ANES respondents.
The comparison includes the variables used for stratification, as well as region, religion, and vote choice in the 2020 U.S. presidential election.
As can be seen in the table, the reduced sample reproduces the weighted composition of the eligible ANES population across the main demographic and political characteristics.

\begin{table}[htbp]
\centering
\caption{\textbf{Demographic and political composition of the reduced ANES sample.}
Comparison between the weighted distribution of eligible respondents in the ANES 2020--2022 Social Media Study and the $N=100$ sample used to construct the personas.
The last column reports the absolute difference in percentage points.}
\label{tab:sample_composition}
\small

\scriptsize
\setlength{\tabcolsep}{4pt}
\renewcommand{\arraystretch}{0.9}

\begin{tabular}{llrrr}
\toprule
Characteristic & Category & Weighted ANES (\%) & $N=100$ & Abs. diff. \\
\midrule

Gender
& Female & 51.7 & 51.0 & 0.7 \\
& Male & 48.3 & 49.0 & 0.7 \\

\addlinespace
Age
& 18--29 & 19.8 & 21.0 & 1.2 \\
& 30--44 & 25.6 & 26.0 & 0.4 \\
& 45--59 & 24.6 & 24.0 & 0.6 \\
& 60--74 & 23.5 & 22.0 & 1.5 \\
& 75+ & 6.4 & 7.0 & 0.6 \\

\addlinespace
Race/ethnicity
& Black & 11.3 & 9.0 & 2.3 \\
& Ethnic minority & 7.6 & 4.0 & 3.6 \\
& Hispanic & 12.2 & 12.0 & 0.2 \\
& White & 68.9 & 75.0 & 6.1 \\

\addlinespace
Education
& Bachelor's degree & 18.2 & 17.0 & 1.2 \\
& High school graduate & 27.8 & 28.0 & 0.2 \\
& Less than high school & 8.9 & 8.0 & 0.9 \\
& Postgraduate & 13.0 & 11.0 & 2.0 \\
& Vocational/technical school & 32.1 & 36.0 & 3.9 \\

\addlinespace
Region
& Midwest & 22.3 & 21.0 & 1.3 \\
& Northeast & 17.6 & 17.0 & 0.6 \\
& South & 37.2 & 35.0 & 2.2 \\
& West & 22.9 & 27.0 & 4.1 \\

\addlinespace
Religion
& Protestant & 26.9 & 22.0 & 4.9 \\
& Roman Catholic & 18.0 & 20.0 & 2.0 \\
& Agnostic & 5.9 & 8.0 & 2.1 \\
& Atheist & 5.0 & 8.0 & 3.0 \\
& Minority religion & 9.1 & 11.0 & 1.9 \\
& Just Christian & 18.9 & 17.0 & 1.9 \\
& Not religious & 16.1 & 14.0 & 2.1 \\

\addlinespace
2020 vote
& Democrat & 43.7 & 43.0 & 0.7 \\
& Independent & 3.0 & 3.0 & 0.0 \\
& Republican & 34.9 & 34.0 & 0.9 \\
& Did not vote & 18.5 & 20.0 & 1.5 \\

\bottomrule
\end{tabular}
\end{table}

\newpage
\section{Network metrics}
\label{sec:network_metrics}
\begin{figure}[ht]
\centering
\includegraphics[width=1\textwidth]{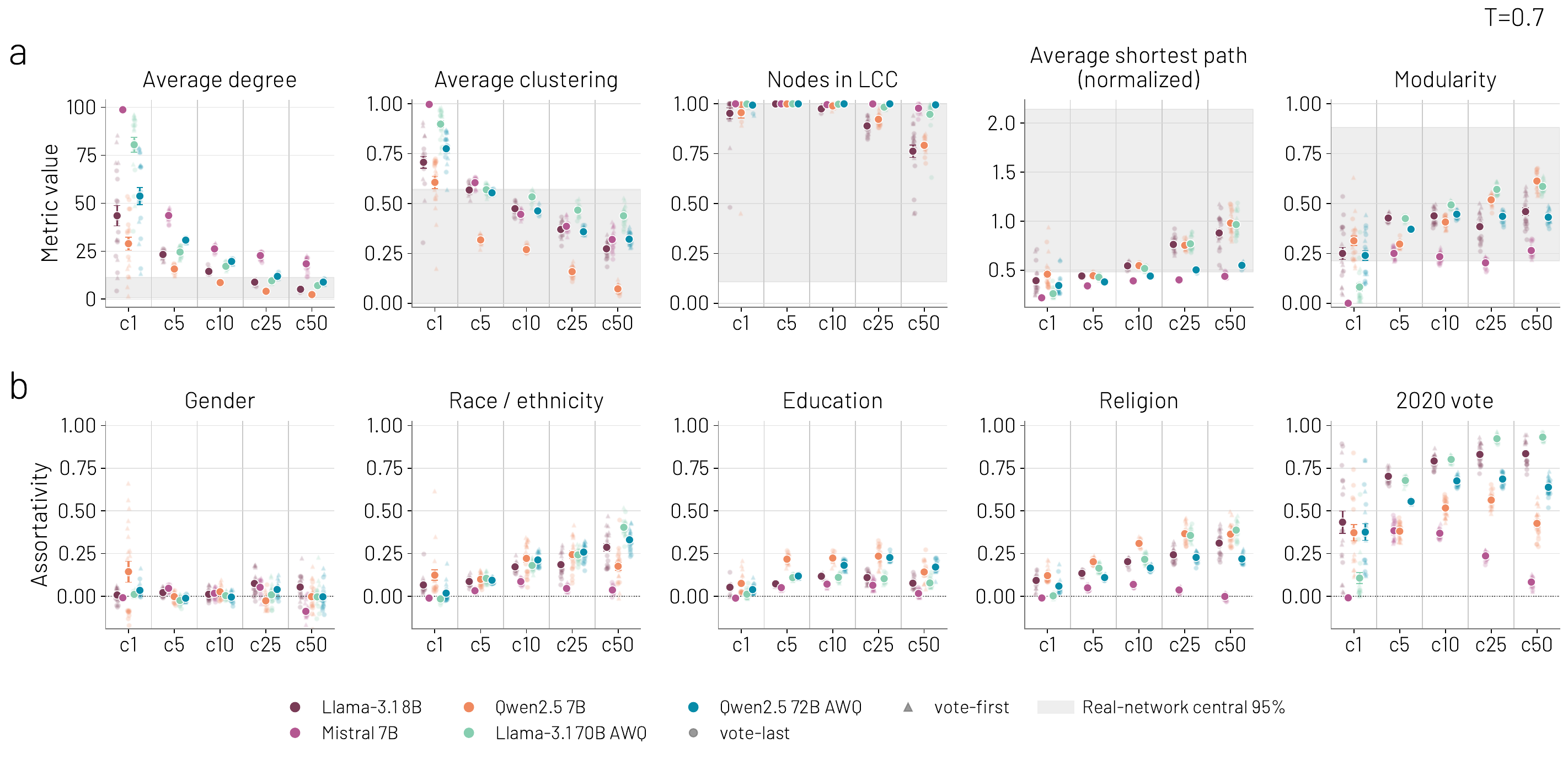}
\caption{\textbf{Network metrics vary across models and prompting techniques for reciprocal networks at $T=0.7$.} The pairwise configuration is indicated by c1, while c5, c10, c25, and c50 indicate the number of candidate alters in each list-based prompt.
Markers represent individual generated networks, with circles and triangles indicating whether vote was positioned last or first in the persona description.
Larger colored points show the models' averages across attribute orders and runs ($10$ for each configuration), and error bars indicate standard errors. The gray areas in panel a represent the central $95\%$ empirical reference ranges derived from the friendship and friendship-like network benchmarks. Panel \textbf{(a)} reports structural network measures, while panel \textbf{(b)} reports assortativity by demographic and political attributes.}\label{SI:metrics07}
\end{figure}

\begin{figure}[ht]
\centering
\includegraphics[width=1\textwidth]{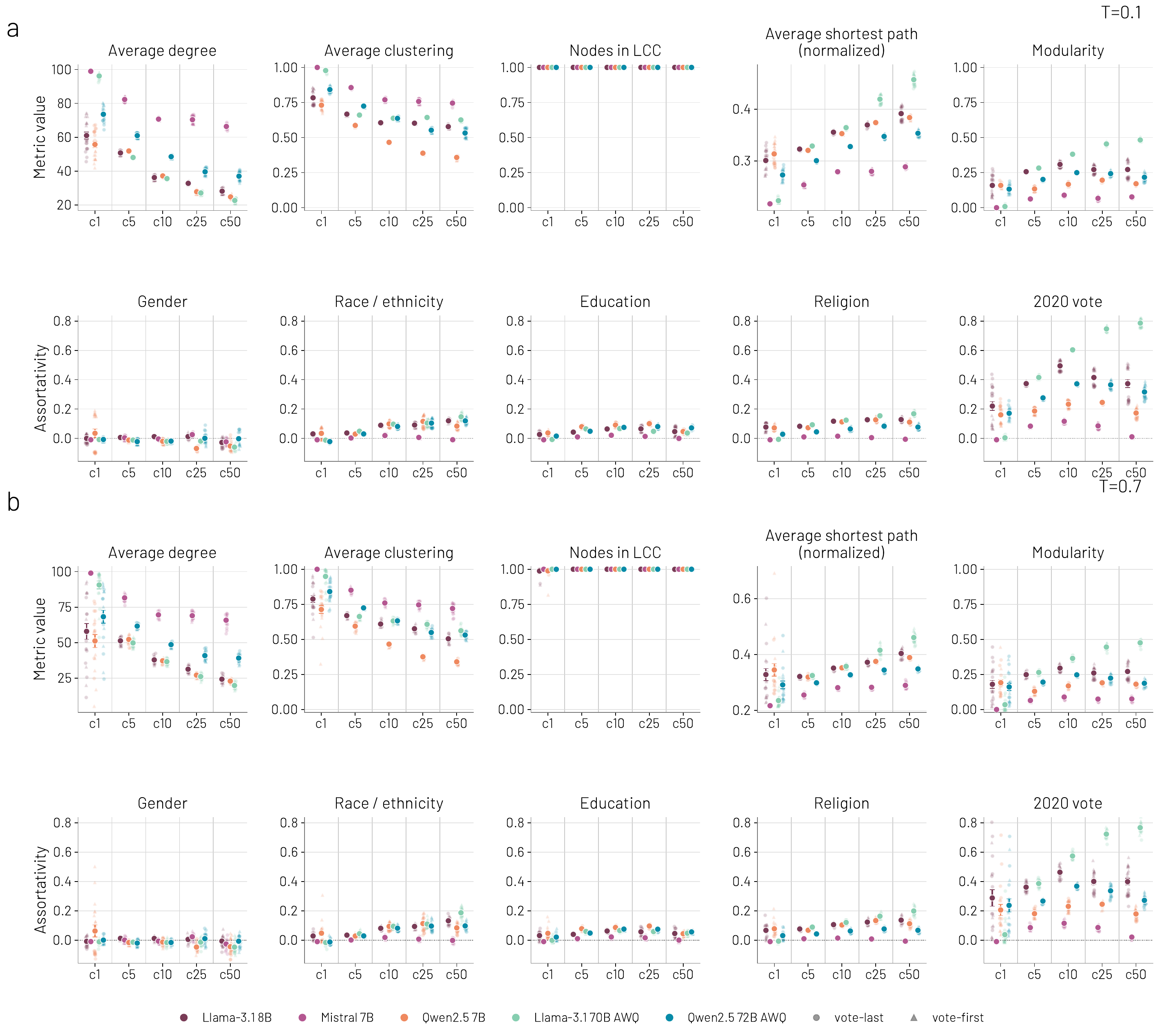}
\caption{\textbf{Network metrics across models and prompting techniques for nomination networks.} Results are shown for $T=0.1$ \textbf{(a)} and $T=0.7$ \textbf{(b)}. Prompt versions contain an increasing number of candidate friends, from one (c1) to fifty (c50), and differ in the ordering of persona attributes, with presidential vote presented first or last. Larger colored points show model averages across attribute orders and runs, and error bars indicate standard errors.}\label{SI:metrics_nomination}
\end{figure}

Fig. \ref{SI:metrics07} reports the network metrics for reciprocal networks at $T=0.7$, complementing Fig. 2 of the main text.
Fig. \ref{SI:metrics_nomination} reports the corresponding measures for the nomination networks at $T=0.1$ and $T=0.7$.
Tables \ref{tab:metrics} and \ref{tab:assortativity_metrics} report the complete structural and assortativity measures for the reciprocal networks, while Tables \ref{tab:metrics_nomination} and \ref{tab:assortativity_metrics_nomination} report the corresponding measures for the nomination networks.
Finally, Table \ref{tab:empirical_networks} describes the empirical friendship and friendship-like networks used as structural benchmarks, and Table \ref{tab:empirical_ranges} reports the resulting empirical reference ranges used to assess structural plausibility.

\clearpage
\begingroup
\scriptsize
\setlength{\tabcolsep}{2.5pt}
\renewcommand{\arraystretch}{0.95}

\begin{longtable}{@{}llllccccc@{}}

\caption{\textbf{Structural properties of the reciprocal LLM-generated networks.}
Values report the mean across 10 runs for each configuration, with standard errors in parentheses.}
\label{tab:metrics}\\

\toprule
Model & $T$ & Method & Order &
Average degree & Clustering & LCC & Normalized path & Modularity \\
\midrule
\endfirsthead

\multicolumn{9}{c}{\textbf{Table \thetable\ continued}}\\
\toprule
Model & $T$ & Method & Order &
Average degree & Clustering & LCC & Normalized path & Modularity \\
\midrule
\endhead

\midrule
\multicolumn{9}{r}{Continued on next page}\\
\endfoot

\bottomrule
\endlastfoot
Llama-3.1 8B & 0.1 & c1 & Vote last & 36.742 (1.127) & 0.717 (0.002) & 1.000 (0.000) & 0.369 (0.006) & 0.342 (0.006) \\
Llama-3.1 8B & 0.1 & c1 & Vote first & 51.538 (1.604) & 0.760 (0.006) & 1.000 (0.000) & 0.324 (0.004) & 0.156 (0.007) \\
Llama-3.1 8B & 0.1 & c5 & Vote last & 24.880 (0.094) & 0.598 (0.001) & 1.000 (0.000) & 0.433 (0.001) & 0.418 (0.001) \\
Llama-3.1 8B & 0.1 & c5 & Vote first & 22.624 (0.103) & 0.557 (0.001) & 1.000 (0.000) & 0.443 (0.002) & 0.451 (0.002) \\
Llama-3.1 8B & 0.1 & c10 & Vote last & 15.118 (0.053) & 0.471 (0.002) & 0.970 (0.003) & 0.536 (0.001) & 0.404 (0.002) \\
Llama-3.1 8B & 0.1 & c10 & Vote first & 14.030 (0.035) & 0.510 (0.002) & 0.987 (0.002) & 0.569 (0.002) & 0.490 (0.001) \\
Llama-3.1 8B & 0.1 & c25 & Vote last & 10.582 (0.069) & 0.422 (0.004) & 0.881 (0.002) & 0.720 (0.004) & 0.326 (0.005) \\
Llama-3.1 8B & 0.1 & c25 & Vote first & 10.968 (0.098) & 0.442 (0.006) & 0.909 (0.003) & 0.701 (0.009) & 0.436 (0.002) \\
\bottomrule
\end{longtable}

\endgroup

\newpage

\begingroup
\scriptsize
\setlength{\tabcolsep}{2.5pt}
\renewcommand{\arraystretch}{0.95}

\begin{longtable}{@{}llllccccc@{}}

\caption{\textbf{Attribute assortativity in the reciprocal LLM-generated networks.}
Values report the mean across 10 runs for each configuration, with standard errors in parentheses.}
\label{tab:assortativity_metrics}\\

\toprule
Model & $T$ & Method & Order &
Gender & Race/ethnicity & Education & Religion & 2020 vote \\
\midrule
\endfirsthead

\multicolumn{9}{c}{\textbf{Table \thetable\ continued}}\\
\toprule
Model & $T$ & Method & Order &
Gender & Race/ethnicity & Education & Religion & 2020 vote \\
\midrule
\endhead

\midrule
\multicolumn{9}{r}{Continued on next page}\\
\endfoot

\bottomrule
\endlastfoot
Llama-3.1 8B & 0.1 & c1 & Vote last & -0.015 (0.001) & 0.050 (0.001) & 0.061 (0.001) & 0.098 (0.002) & 0.578 (0.016) \\
Llama-3.1 8B & 0.1 & c1 & Vote first & 0.044 (0.001) & 0.074 (0.001) & 0.049 (0.003) & 0.099 (0.004) & 0.215 (0.015) \\
Llama-3.1 8B & 0.1 & c5 & Vote last & 0.016 (0.001) & 0.061 (0.001) & 0.073 (0.001) & 0.136 (0.001) & 0.690 (0.001) \\
Llama-3.1 8B & 0.1 & c5 & Vote first & 0.041 (0.001) & 0.102 (0.001) & 0.075 (0.001) & 0.144 (0.001) & 0.734 (0.003) \\
Llama-3.1 8B & 0.1 & c10 & Vote last & -0.022 (0.003) & 0.183 (0.001) & 0.117 (0.001) & 0.198 (0.001) & 0.770 (0.002) \\
Llama-3.1 8B & 0.1 & c10 & Vote first & 0.015 (0.003) & 0.188 (0.001) & 0.140 (0.002) & 0.222 (0.001) & 0.838 (0.002) \\
Llama-3.1 8B & 0.1 & c25 & Vote last & 0.035 (0.006) & 0.126 (0.003) & 0.124 (0.002) & 0.190 (0.003) & 0.801 (0.002) \\
Llama-3.1 8B & 0.1 & c25 & Vote first & 0.088 (0.007) & 0.197 (0.004) & 0.074 (0.003) & 0.251 (0.002) & 0.917 (0.003) \\
\bottomrule
\end{longtable}

\endgroup

\newpage

\begingroup
\scriptsize
\setlength{\tabcolsep}{2.5pt}
\renewcommand{\arraystretch}{0.95}

\begin{longtable}{@{}llllccccc@{}}

\caption{\textbf{Structural properties of the nomination LLM-generated networks.}
Values report the mean across 10 runs for each configuration, with standard errors in parentheses.}
\label{tab:metrics_nomination}\\

\toprule
Model & $T$ & Method & Order &
Average degree & Clustering & LCC & Normalized path & Modularity \\
\midrule
\endfirsthead

\multicolumn{9}{c}{\textbf{Table \thetable\ continued}}\\
\toprule
Model & $T$ & Method & Order &
Average degree & Clustering & LCC & Normalized path & Modularity \\
\midrule
\endhead

\midrule
\multicolumn{9}{r}{Continued on next page}\\
\endfoot

\bottomrule
\endlastfoot
Llama-3.1 8B & 0.1 & c1 & Vote last & 53.338 (1.311) & 0.733 (0.002) & 1.000 (0.000) & 0.318 (0.003) & 0.225 (0.009) \\
Llama-3.1 8B & 0.1 & c1 & Vote first & 68.656 (1.606) & 0.834 (0.006) & 1.000 (0.000) & 0.284 (0.004) & 0.092 (0.005) \\
Llama-3.1 8B & 0.1 & c5 & Vote last & 51.862 (0.124) & 0.674 (0.001) & 1.000 (0.000) & 0.321 (0.000) & 0.255 (0.001) \\
Llama-3.1 8B & 0.1 & c5 & Vote first & 49.616 (0.131) & 0.659 (0.001) & 1.000 (0.000) & 0.325 (0.000) & 0.258 (0.001) \\
Llama-3.1 8B & 0.1 & c10 & Vote last & 37.422 (0.095) & 0.606 (0.001) & 1.000 (0.000) & 0.352 (0.000) & 0.286 (0.001) \\
Llama-3.1 8B & 0.1 & c10 & Vote first & 34.936 (0.090) & 0.605 (0.001) & 1.000 (0.000) & 0.359 (0.000) & 0.330 (0.001) \\
Llama-3.1 8B & 0.1 & c25 & Vote last & 32.994 (0.134) & 0.599 (0.001) & 1.000 (0.000) & 0.366 (0.001) & 0.240 (0.002) \\
Llama-3.1 8B & 0.1 & c25 & Vote first & 32.506 (0.121) & 0.606 (0.001) & 1.000 (0.000) & 0.373 (0.001) & 0.304 (0.002) \\
\bottomrule
\end{longtable}

\endgroup

\newpage

\begingroup
\scriptsize
\setlength{\tabcolsep}{2.5pt}
\renewcommand{\arraystretch}{0.95}

\begin{longtable}{@{}llllccccc@{}}

\caption{\textbf{Attribute assortativity in the nomination LLM-generated networks.}
Values report the mean across 10 runs for each configuration, with standard errors in parentheses.}
\label{tab:assortativity_metrics_nomination}\\

\toprule
Model & $T$ & Method & Order &
Gender & Race/ethnicity & Education & Religion & 2020 vote \\
\midrule
\endfirsthead

\multicolumn{9}{c}{\textbf{Table \thetable\ continued}}\\
\toprule
Model & $T$ & Method & Order &
Gender & Race/ethnicity & Education & Religion & 2020 vote \\
\midrule
\endhead

\midrule
\multicolumn{9}{r}{Continued on next page}\\
\endfoot

\bottomrule
\endlastfoot
Llama-3.1 8B & 0.1 & c1 & Vote last & -0.033 (0.002) & 0.022 (0.001) & 0.039 (0.002) & 0.100 (0.001) & 0.345 (0.015) \\
Llama-3.1 8B & 0.1 & c1 & Vote first & 0.031 (0.000) & 0.038 (0.001) & 0.013 (0.002) & 0.053 (0.004) & 0.094 (0.010) \\
Llama-3.1 8B & 0.1 & c5 & Vote last & 0.010 (0.001) & 0.034 (0.000) & 0.043 (0.001) & 0.074 (0.001) & 0.362 (0.002) \\
Llama-3.1 8B & 0.1 & c5 & Vote first & 0.001 (0.001) & 0.039 (0.001) & 0.041 (0.001) & 0.090 (0.001) & 0.384 (0.002) \\
Llama-3.1 8B & 0.1 & c10 & Vote last & 0.008 (0.001) & 0.082 (0.001) & 0.066 (0.001) & 0.116 (0.001) & 0.457 (0.001) \\
Llama-3.1 8B & 0.1 & c10 & Vote first & 0.015 (0.001) & 0.095 (0.002) & 0.061 (0.001) & 0.117 (0.001) & 0.532 (0.002) \\
Llama-3.1 8B & 0.1 & c25 & Vote last & 0.001 (0.003) & 0.104 (0.002) & 0.082 (0.001) & 0.123 (0.001) & 0.358 (0.002) \\
Llama-3.1 8B & 0.1 & c25 & Vote first & 0.023 (0.003) & 0.077 (0.001) & 0.047 (0.001) & 0.130 (0.001) & 0.472 (0.002) \\
\bottomrule
\end{longtable}

\endgroup

\newpage

\begingroup
\scriptsize
\setlength{\tabcolsep}{2.5pt}
\renewcommand{\arraystretch}{1.05}

\begin{longtable}{@{}p{4.0cm}ccp{3.0cm}p{4.0cm}@{}}

\caption{\textbf{Empirical friendship and friendship-like networks used to define the structural reference ranges.}
The benchmark contains 57 networks from 13 datasets. For datasets containing multiple networks, the number of networks and range of network sizes are reported. Values of $N$ are derived from the graph objects used in the analysis; the High School 2013 graph is edge-induced, and the MSW Students graphs retain the full 145-student roster. Network-level observations are weighted so that each dataset contributes equal total weight when estimating the empirical reference ranges.}
\label{tab:empirical_networks}\\

\toprule
Dataset & No. networks & $N$ & Relation & Reference \\
\midrule
\endfirsthead

\multicolumn{5}{c}{\textbf{Table \thetable\ continued}}\\
\toprule
Dataset & No. networks & $N$ & Relation & Reference \\
\midrule
\endhead

\midrule
\multicolumn{5}{r}{Continued on next page}\\
\endfoot

\bottomrule
\endlastfoot
Baerveldt Dutch Social Behavior & 38 & 31--91 & Emotional support & \citet{houtzager_just_1999}\\
Galesburg & 1 & 31 & Friendship & \citet{coleman_diffusion_1957} \\
Hi-tech & 1 & 36 & Friendship & \citet{krackhardt1999} \\
High School 2013 & 1 & 134 & Friendship & \citet{mastrandrea_contact_2015}\\
Tailor & 2 & 39 & Sociational & \citet{kapferer_strategy_1972}\\
Knecht class 12b & 4 & 26 & Friendship & \citet{knecht_friendship_2010}\\
Lazega Lawyers & 2 & 36--71 & Friendship & \citet{lazega_collegial_2001}\\
MSW Students & 3 & 145 & Friendship & \citet{mauldin_relationships_2017}\\
Moreno High School & 1 & 70 & Friendship & \citet{rossi_network_2015}\\
PEAS & 1 & 213 & Friendship & \citet{oosterhoff_early_2022}\\
Prison & 1 & 67 & Friendship & \citet{macrae_direct_1960}\\
Sciences Po & 1 & 543 & Friendship & \citet{algan_friendship_2026}\\
Karate & 1 & 34 & Social ties & \citet{zachary_information_1977}\\
\midrule
\textbf{Total} & \textbf{57} & & & \\
\bottomrule
\end{longtable}

\endgroup

\newpage

\begin{table}[ht]
\centering
\caption{\textbf{Empirical reference ranges used to assess structural plausibility.}
Ranges correspond to the weighted 2.5th and 97.5th percentiles across the 57 empirical networks, with equal total weight assigned to each of the 13 datasets.}
\label{tab:empirical_ranges}
\begin{tabular}{lcc}
\toprule
Metric & 2.5th percentile & 97.5th percentile \\
\midrule
Average degree & 0.679 & 11.436 \\
Average clustering & 0.000 & 0.571 \\
Nodes in LCC & 0.106 & 1.000 \\
Normalized shortest path & 0.484 & 2.141 \\
Modularity & 0.212 & 0.881 \\
\bottomrule
\end{tabular}
\end{table}

\newpage
\section{Partisan exposure results}
\label{sec:exposure}
\begin{figure}[ht]
\centering
\includegraphics[width=1\textwidth]{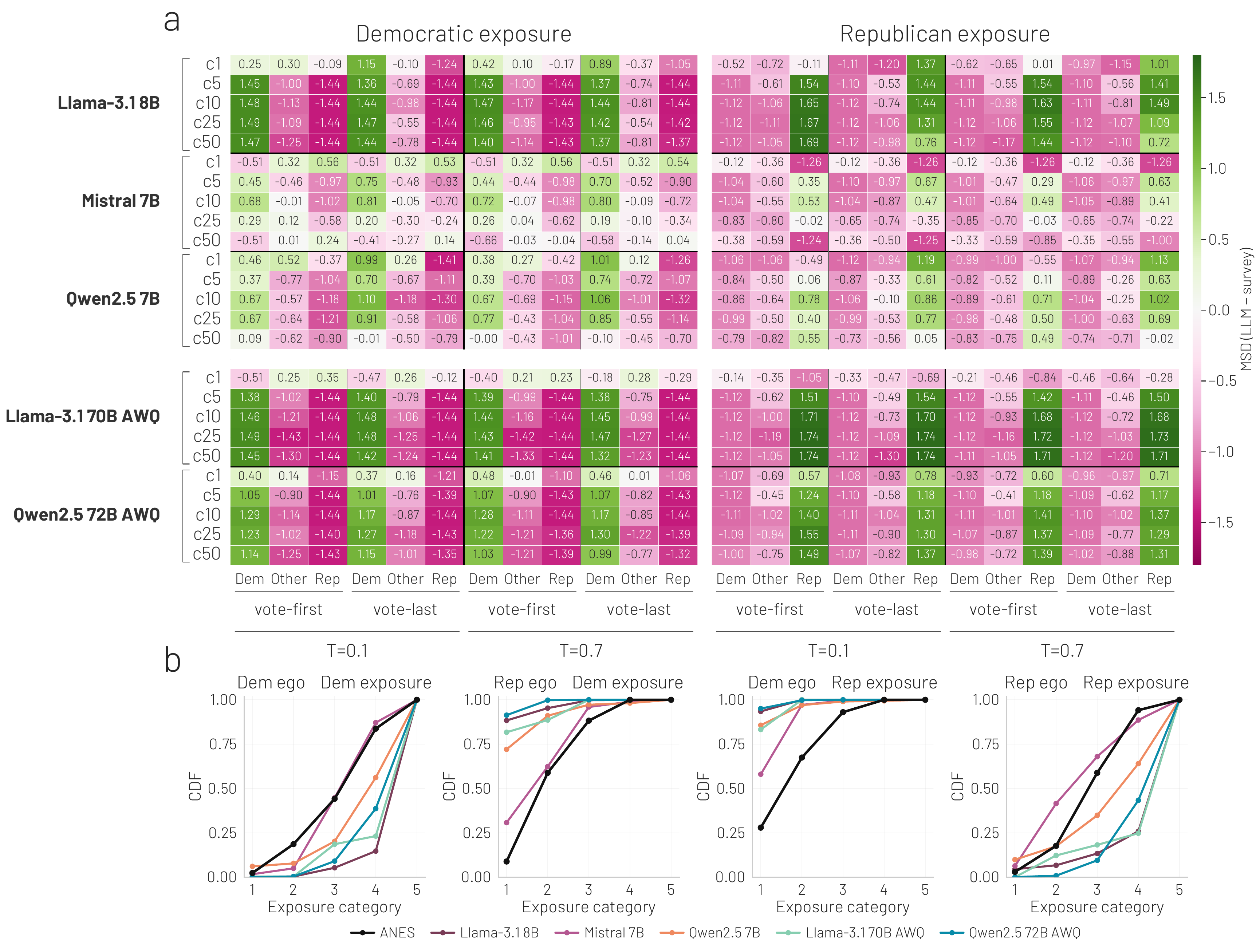}
\caption{\textbf{Distance between partisan exposure in LLM-generated networks and ANES survey using reciprocal networks.} Panel \textbf{(a)} reports the Mean Signed Deviation (MSD) between generated and survey-reported exposure across models and experimental configurations. Positive values indicate overestimation relative to the ANES survey, while negative values indicate underestimation. Panel \textbf{(b)} reports the corresponding cumulative distributions of partisan exposure for Democratic and Republican egos.}
\label{SI:exposure_reciprocal}
\end{figure}

Fig. \ref{SI:exposure_reciprocal} reports the corresponding partisan-exposure results for the reciprocal networks.
Table \ref{tab:complete_msd_nomination} reports the complete Mean Signed Deviation (MSD) values between LLM-generated and survey-reported partisan exposure for the nomination networks across models, prompting methods, temperatures, and attribute orderings.
Table \ref{tab:complete_msd_reciprocal} reports the complete MSD values for the reciprocal networks.

\newpage

\begin{longtable}{@{}lllllcc@{}}
\caption{\textbf{Mean Signed Deviation between LLM-generated and survey-reported partisan exposure in nomination networks.} Values report the unweighted mean across 10 runs for each experimental configuration, with standard errors in parentheses. Positive values indicate overestimation relative to the corresponding ANES personas, while negative values indicate underestimation. Order indicates whether vote was presented first or last in the persona description.} \label{tab:complete_msd_nomination} \\
\toprule
Model & T & Method & Order & Ego & Dem. exposure & Rep. exposure \\
\midrule
\endfirsthead
\caption[]{\textbf{Mean Signed Deviation between LLM-generated and survey-reported partisan exposure in nomination networks.} Values report the unweighted mean across 10 runs for each experimental configuration, with standard errors in parentheses. Positive values indicate overestimation relative to the corresponding ANES personas, while negative values indicate underestimation. Order indicates whether vote was presented first or last in the persona description.} \\
\toprule
Model & T & Method & Order & Ego & Dem. exposure & Rep. exposure \\
\midrule
\endhead
\midrule
\multicolumn{7}{r}{Continued on next page} \\
\midrule
\endfoot
\bottomrule
\endlastfoot
Llama-3.1 8B & 0.1 & c1 & First & Democratic & -0.167 (0.048) & -0.316 (0.025) \\
Llama-3.1 8B & 0.1 & c1 & First & Republican & -0.126 (0.045) & -0.335 (0.037) \\
Llama-3.1 8B & 0.1 & c1 & First & Other & 0.318 (0.015) & -0.723 (0.005) \\
Llama-3.1 8B & 0.1 & c1 & Last & Democratic & 0.588 (0.058) & -1.102 (0.006) \\
Llama-3.1 8B & 0.1 & c1 & Last & Republican & -1.144 (0.032) & 0.591 (0.065) \\
Llama-3.1 8B & 0.1 & c1 & Last & Other & -0.241 (0.034) & -1.245 (0.007) \\
Llama-3.1 8B & 0.1 & c5 & First & Democratic & 1.449 (0.010) & -1.116 (0.000) \\
Llama-3.1 8B & 0.1 & c5 & First & Republican & -1.250 (0.008) & 0.712 (0.004) \\
\end{longtable}

\newpage

\begin{longtable}{@{}lllllcc@{}}
\caption{\textbf{Mean Signed Deviation between LLM-generated and survey-reported partisan exposure in reciprocal networks.} Values report the unweighted mean across 10 runs for each experimental configuration, with standard errors in parentheses. Positive values indicate overestimation relative to the corresponding ANES personas, while negative values indicate underestimation. Order indicates whether vote was presented first or last in the persona description.} \label{tab:complete_msd_reciprocal} \\
\toprule
Model & T & Method & Order & Ego & Dem. exposure & Rep. exposure \\
\midrule
\endfirsthead
\caption[]{\textbf{Mean Signed Deviation between LLM-generated and survey-reported partisan exposure in reciprocal networks.} Values report the unweighted mean across 10 runs for each experimental configuration, with standard errors in parentheses. Positive values indicate overestimation relative to the corresponding ANES personas, while negative values indicate underestimation. Order indicates whether vote was presented first or last in the persona description.} \\
\toprule
Model & T & Method & Order & Ego & Dem. exposure & Rep. exposure \\
\midrule
\endhead
\midrule
\multicolumn{7}{r}{Continued on next page} \\
\midrule
\endfoot
\bottomrule
\endlastfoot
Llama-3.1 8B & 0.1 & c1 & First & Democratic & 0.247 (0.067) & -0.516 (0.034) \\
Llama-3.1 8B & 0.1 & c1 & First & Republican & -0.091 (0.035) & -0.106 (0.043) \\
Llama-3.1 8B & 0.1 & c1 & First & Other & 0.305 (0.021) & -0.718 (0.034) \\
Llama-3.1 8B & 0.1 & c1 & Last & Democratic & 1.153 (0.028) & -1.112 (0.003) \\
Llama-3.1 8B & 0.1 & c1 & Last & Republican & -1.244 (0.016) & 1.371 (0.036) \\
Llama-3.1 8B & 0.1 & c1 & Last & Other & -0.105 (0.052) & -1.200 (0.012) \\
Llama-3.1 8B & 0.1 & c5 & First & Democratic & 1.449 (0.006) & -1.107 (0.004) \\
Llama-3.1 8B & 0.1 & c5 & First & Republican & -1.441 (0.000) & 1.538 (0.004) \\
\end{longtable}

\newpage
\section{Political assortativity and partisan exposure}
\label{sec:assortativity_exposure}
\begin{figure}[h!]
\centering
\includegraphics[width=1\textwidth]{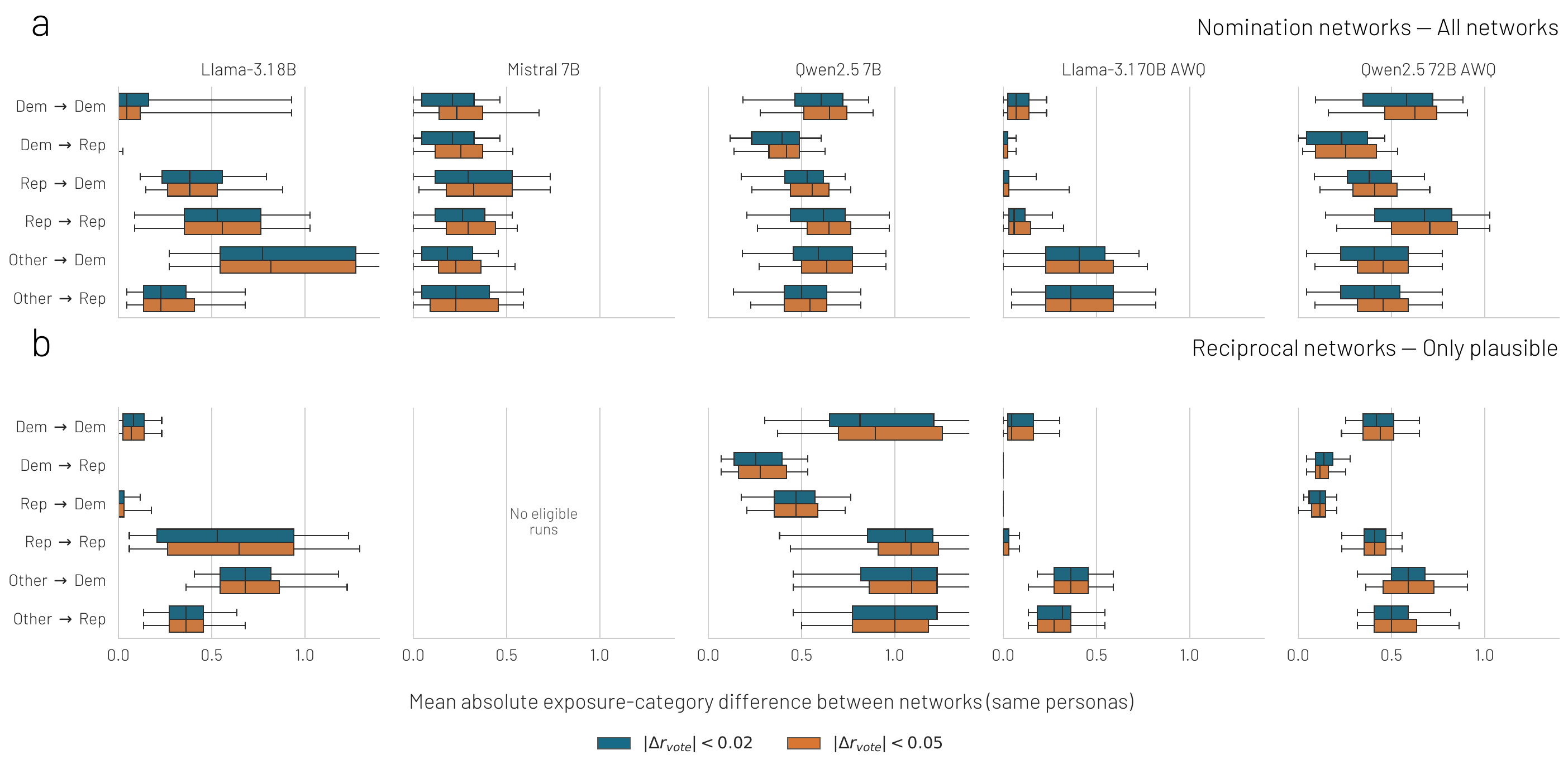}
\caption{\textbf{Mean absolute within-person differences in partisan exposure between pairs of LLM-generated networks from the same model whose vote assortativity differs by less than 0.02 and 0.05.} Results are shown for nomination networks \textbf{(a)} and for reciprocal networks meeting all five structural plausibility criteria \textbf{(b)}. For each pair, exposure is compared for the same personas on the 1–5 exposure scale. Boxes show the interquartile range, center lines indicate medians, and whiskers the 5th–95th percentiles.}\label{SI:assortativity_matching}
\end{figure}

Fig.~\ref{SI:assortativity_matching} repeats the matching analysis described in the main text using a less restrictive assortativity threshold of $0.05$ and compares the results with those obtained using the main threshold of $0.02$. 
Results are shown for the nomination networks and for the reciprocal networks that satisfy all five structural plausibility criteria. 
Table~\ref{tab:matched_pairs} reports the number of eligible matched network pairs for both thresholds and both network representations.

\begin{table}[htbp]
\centering
\caption{\textbf{Number of matched network pairs with similar political assortativity.}
Number of eligible pairs of LLM-generated networks from the same model under the two
assortativity thresholds used in the matching analysis. Nomination-network counts use
all generated networks, whereas reciprocal-network counts use only networks meeting all
five structural plausibility criteria. Matched networks differ in experimental
configuration.}
\label{tab:matched_pairs}
\begin{tabular}{@{}lrrrr@{}}
\toprule
& \multicolumn{2}{c}{Nomination networks---all} &
  \multicolumn{2}{c}{Reciprocal networks---plausible} \\
\cmidrule(lr){2-3}\cmidrule(l){4-5}
Model & $|\Delta r_{\mathrm{vote}}|<0.02$ & $|\Delta r_{\mathrm{vote}}|<0.05$ &
        $|\Delta r_{\mathrm{vote}}|<0.02$ & $|\Delta r_{\mathrm{vote}}|<0.05$ \\
\midrule
Llama-3.1 8B      & 3,092 & 4,652 &   484 & 1,100 \\
Mistral 7B        & 2,804 & 6,534 &     0 &     0 \\
Qwen2.5 7B        & 4,221 & 7,818 & 1,019 & 2,352 \\
Llama-3.1 70B AWQ & 3,039 & 3,729 & 1,723 & 2,767 \\
Qwen2.5 72B AWQ   & 4,300 & 6,124 &   166 &   415 \\
\bottomrule
\end{tabular}
\end{table}

\clearpage
\section{Group-specific exposure results}
\label{sec:groups}
\begin{figure}[h!]
\centering
\includegraphics[width=1\textwidth]{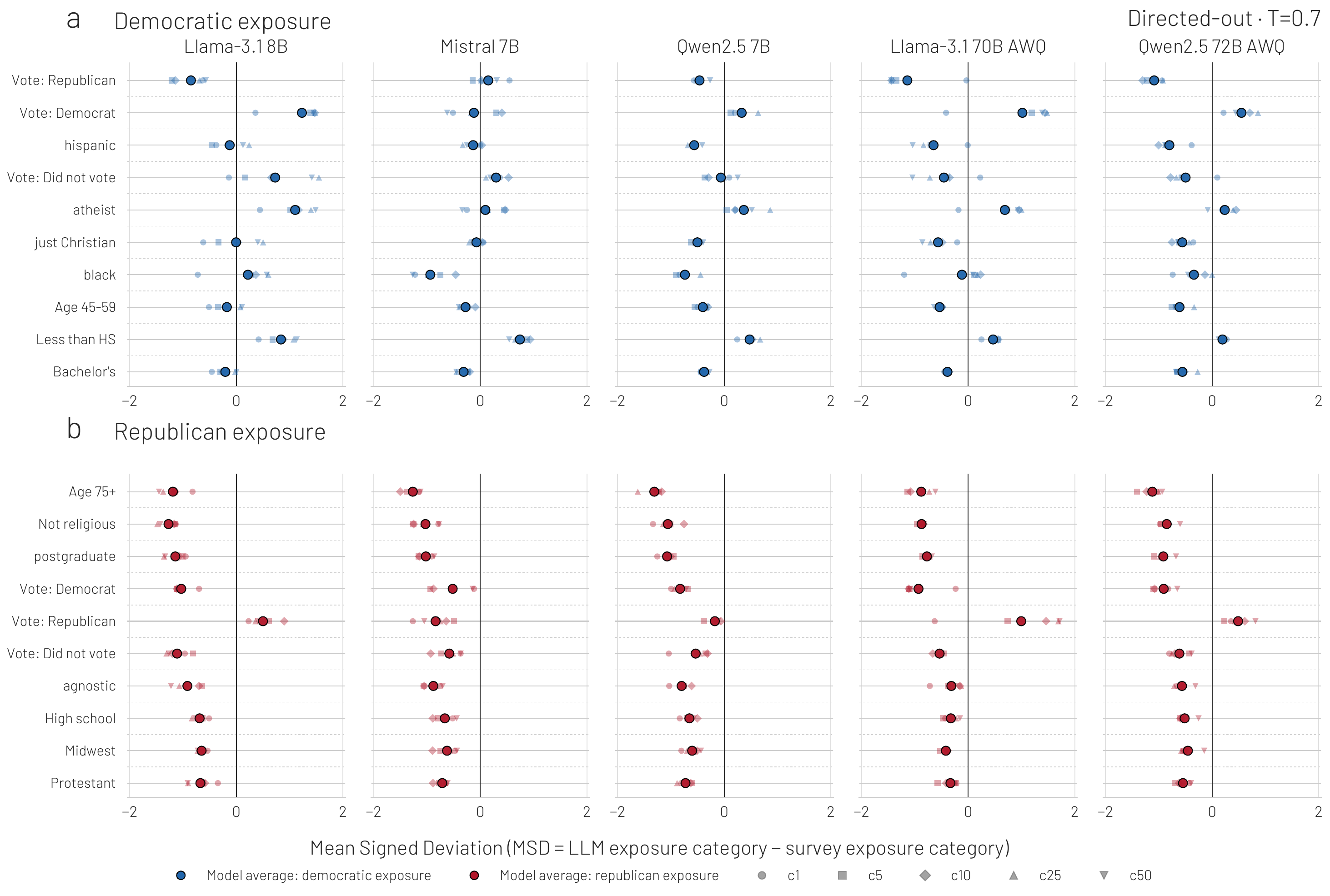}
\caption{{\textbf{Group-specific errors in partisan exposure in LLM-generated networks at $T=0.7$.} Mean Signed Deviation (MSD) between LLM-generated and survey-reported exposure to \textbf{(a)} Democrats and \textbf{(b)} Republicans for the social groups with the largest deviations. Negative values indicate underestimation relative to the survey, while positive values indicate overestimation. Different point styles represent the different prompting methods, with larger markers showing the model average.} }\label{SI:group_specific_T07}
\end{figure}

Fig.~\ref{SI:group_specific_T07} reports the group-specific Mean Signed Deviation (MSD) in partisan exposure at $T=0.7$, complementing Fig. 5 of the main text, which reports the corresponding results at $T=0.1$, while Table \ref{tab:group_specific_exposure_errors} shows the complete set of group-specific MSD values for all demographic and political groups represented by at least five personas in the sampled population.

\begingroup
\scriptsize
\setlength{\tabcolsep}{3pt}
\renewcommand{\arraystretch}{0.95}
\begin{longtable}{@{}lllp{2.8cm}rrr@{}}
\caption{\textbf{Complete group-specific errors in LLM-generated partisan exposure.} Mean Signed Deviation (MSD) between LLM-generated and survey-reported exposure to Democrats and Republicans in nomination networks for all demographic and political groups with at least five personas. Values report model averages across prompting methods and attribute orderings separately for each temperature, with standard errors across configurations in parentheses. Positive values indicate overestimation relative to the ANES survey, while negative values indicate underestimation.}\label{tab:group_specific_exposure_errors}\\
\toprule
Model & $T$ & Attribute & Group & $N$ & Dem. exposure & Rep. exposure \\
\midrule
\endfirsthead
\multicolumn{7}{c}{\textbf{Table \thetable\ continued}}\\
\toprule
Model & $T$ & Attribute & Group & $N$ & Dem. exposure & Rep. exposure \\
\midrule
\endhead
\midrule
\multicolumn{7}{r}{Continued on next page}\\
\endfoot
\bottomrule
\endlastfoot
Llama-3.1 8B & 0.1 & Race/ethnicity & black & 9 & 0.224 (0.176) & -0.109 (0.079) \\
Llama-3.1 8B & 0.7 & Race/ethnicity & black & 9 & 0.218 (0.163) & -0.093 (0.057) \\
Mistral 7B & 0.1 & Race/ethnicity & black & 9 & -0.953 (0.107) & 0.477 (0.099) \\
Mistral 7B & 0.7 & Race/ethnicity & black & 9 & -0.937 (0.106) & 0.469 (0.094) \\
Qwen2.5 7B & 0.1 & Race/ethnicity & black & 9 & -0.719 (0.081) & 0.197 (0.063) \\
Qwen2.5 7B & 0.7 & Race/ethnicity & black & 9 & -0.737 (0.069) & 0.212 (0.044) \\
Llama-3.1 70B AWQ & 0.1 & Race/ethnicity & black & 9 & -0.117 (0.187) & 0.101 (0.117) \\
Llama-3.1 70B AWQ & 0.7 & Race/ethnicity & black & 9 & -0.117 (0.183) & 0.080 (0.111) \\
\end{longtable}
\endgroup

\bibliography{sn-bibliography}